\documentclass[twocolumn,floatfix,superscriptaddress,amsmath,amssymb,aps,pra]{revtex4-2}

\pdfoutput=1
\usepackage{graphicx}
\graphicspath{ {Figures/} }

\usepackage{dcolumn}
\usepackage{bm}
\usepackage{bbold}

\usepackage{pgfplots, pgfplotstable}
\usepgfplotslibrary{groupplots}
\pgfplotsset{compat=1.18}

\usetikzlibrary{shapes.misc}
\pgfplotsset{
  colormap/viridis_r/.style={
    colormap={viridis_r}{
      rgb255(0cm)=(253,231,36)
      rgb255(1cm)=(181,221,43)
      rgb255(2cm)=(109,206,88)
      rgb255(3cm)=(53,183,120)
      rgb255(4cm)=(30,157,136)
      rgb255(5cm)=(37,130,142)
      rgb255(6cm)=(48,103,141)
      rgb255(7cm)=(62,73,137)
      rgb255(8cm)=(71,39,119)
      rgb255(9cm)=(68,1,84)
    }
  }
}
\pgfplotsset{
  colormap/iridescent/.style={
    colormap={iridescent}{
      rgb255(0cm)=(254,251,233)
      rgb255(1cm)=(252,247,213)
      rgb255(2cm)=(245,243,193)
      rgb255(3cm)=(234,240,181)
      rgb255(4cm)=(221,236,191)
      rgb255(5cm)=(208,231,202)
      rgb255(6cm)=(194,227,210)
      rgb255(7cm)=(181,221,216)
      rgb255(8cm)=(168,216,220)
      rgb255(9cm)=(155,210,225)
      rgb255(10cm)=(141,203,228)
      rgb255(11cm)=(129,196,231)
      rgb255(12cm)=(123,188,231)
      rgb255(13cm)=(126,178,228)
      rgb255(14cm)=(136,165,221)
      rgb255(15cm)=(147,152,210)
      rgb255(16cm)=(155,138,196)
      rgb255(17cm)=(157,125,178)
      rgb255(18cm)=(154,112,158)
      rgb255(19cm)=(144,99,136)
      rgb255(20cm)=(128,87,112)
      rgb255(21cm)=(104,73,87)
      rgb255(22cm)=(70,53,58)
    }
  }
}
\pgfplotsset{
  colormap/ylorbr/.style={
    colormap={ylorbr}{
      rgb255(0cm)=(255,255,229)
      rgb255(1cm)=(255,247,188)
      rgb255(2cm)=(254,227,145)
      rgb255(3cm)=(254,196,79)
      rgb255(4cm)=(251,154,41)
      rgb255(5cm)=(236,112,20)
      rgb255(6cm)=(204,76,2)
      rgb255(7cm)=(153,52,4)
      rgb255(8cm)=(102,37,6)
    }
  }
}
\definecolor{tolblue}{HTML}{4477AA}
\definecolor{tolred}{HTML}{EE6677}
\definecolor{tolgreen}{HTML}{228833}

\usepackage{adjustbox}
\usepackage{yquant}
    \yquantset{register/default name=$\reg_{\idx}$}
    \yquantset{register/minimum height=14pt}
    \yquantset{operator/separation=6pt}
    \yquantset{every circuit/.style={line width=1.0pt}}
    \yquantset{every operator/.style={scale=1.2}}
    \yquantset{every control/.style={shape=yquant-circle, anchor=center, radius=.75mm}}

\usepackage{multirow}
\usepackage{booktabs}

\usepackage{pgfplots}
\usepackage{pgfplotstable}
\usepgfplotslibrary{fillbetween}
\pgfplotsset{compat=1.18}

\usepackage{array,colortbl,xcolor}

\usepackage[utf8]{inputenc}
\usepackage[T1]{fontenc}
\usepackage{mathrsfs,bbm}
\usepackage{dsfont}

\usepackage{amssymb}
\usepackage{amsmath}
\usepackage{amsfonts}
\usepackage{amsthm}
\usepackage{mathtools}
\usepackage{hyperref}
\hypersetup{pdfpagemode=UseNone}
\hypersetup{colorlinks=true}
\hypersetup{citecolor=blue}
\hypersetup{linkcolor=blue}
\hypersetup{urlcolor=blue}
\usepackage{cleveref}

\usepackage{latexsym}
\usepackage{color}
\usepackage{enumerate}
\usepackage{comment}

\newtheorem{proposition}{Proposition}
\newtheorem{theorem}[proposition]{Theorem}
\newtheorem{lemma}[proposition]{Lemma}
\newtheorem{corollary}{Corollary}
\newtheorem{remark}{Remark}

\theoremstyle{definition}

\newcommand{\mo}[1]{\left| #1 \right|} 

\newcommand{\ip}[2]{\langle #1 | #2 \rangle} 
\newcommand{\ket}[1]{|#1\rangle} 
\newcommand{\bra}[1]{\langle#1|} 
\newcommand{\no}[1]{\left\|#1\right\|} 
\newcommand{\id}{\mathbbm{1}} 

\newcommand{\ceil}[1]{\lceil {#1} \rceil}

\date{\today}

\begin{document}

\title{Minimizing entanglement entropy for enhanced quantum state preparation}

\author{Oskari Kerppo}
\email{oskari.kerppo@quanscient.com}
\affiliation{Quanscient Oy, Peltokatu 34, 33100 Tampere, Finland}

\author{William Steadman}
\affiliation{Quanscient Oy, Peltokatu 34, 33100 Tampere, Finland}

\author{Ossi Niemimäki}
\affiliation{Quanscient Oy, Peltokatu 34, 33100 Tampere, Finland}

\author{Valtteri Lahtinen}
\affiliation{Quanscient Oy, Peltokatu 34, 33100 Tampere, Finland}
\affiliation{School of Engineering Science, Lappeenranta–Lahti University of Technology, P.O. Box 20, 53851, Lappeenranta, Finland}

\begin{abstract}
Quantum state preparation is an important subroutine in many quantum algorithms. The goal is to encode classical information directly to the quantum state so that it is possible to leverage quantum algorithms for data processing. However, quantum state preparation of arbitrary states scales exponentially in the number of two-qubit gates, and this makes quantum state preparation a very difficult task on quantum computers, especially on near-term noisy devices. This represents a major challenge in achieving quantum advantage. We present and analyze a novel two-step state preparation method where we first minimize the entanglement entropy of the target quantum state, thus transforming the state to one that is easier to prepare. The state with reduced entanglement entropy is then represented as a matrix product state, resulting in a high accuracy preparation of the target state. Our method is suitable for NISQ devices and we give rigorous lower bounds on the accuracy of the prepared state in terms of the entanglement entropy. We benchmark our method with 2D normal distribution and Ricker wavelet states with 6--20 qubits.
\end{abstract}

\maketitle

\section{Introduction}

A quantum computer is typically thought of as a device consisting of $n$ qubits initially in an unentangled state $\ket{0}$. The goal of quantum computing is to manipulate the state of these qubits in a clever way so that advanced information processing tasks become possible. In quantum computing such tasks are typically called quantum algorithms, famous examples being Shor's algorithm \cite{Shor1997Polynomial} for integer factorization, Grover's search algorithm \cite{Grover1996Fast} and HHL algorithm for solving linear systems of equations \cite{Harrow2009Quantum}. 

The first step in many quantum algorithms is quantum state preparation (QSP) \cite{Shende2006Synthesis, Plesch2011Quantum, Iten2016Quantum} where the initial state $\ket{0}$ is transformed into some other state $\ket{\psi}$ before the actual algorithm is executed. It is possible to encode classical data into the complex amplitudes of the state $\ket{\psi}$: \begin{align}
    \ket{\psi} = \sum_{i=1}^{2^N} \alpha_i \ket{i},
\end{align}
where $\alpha_i$'s are complex numbers with $\sum_i \mo{\alpha_i}^2 = 1$ and $\{ \ket{i} \}_{i=1}^{2^N}$ is the computational basis for $N$ qubits. One could, for instance, encode image data into the state $\ket{\psi}$ for advanced quantum image processing \cite{Haque2023}. Throughout this work we assume the amplitudes to be real-valued.

A key performance metric for quantum algorithms is how the circuit depth and quantum gate count, especially the number of 2-qubit gates, scales when the number of qubits grows. A simple parameter counting argument shows that general algorithms for preparing an arbitrary quantum state require at least $2^{N-1}$ CX gates \cite{Plesch2011Quantum, Iten2016Quantum}. Thus, generally speaking, QSP scales exponentially in the number of qubits and is unfeasible for a large numbers of qubits. This presents a major obstacle for reaching quantum advantage, as even if efficient quantum algorithms exist for data processing, it might be infeasible to initialize the quantum computer in the desired state. However, the lower bound on CX gates for QSP is only valid for unstructured data. Many quantum states allow efficient preparation methods, and the research community is constantly pushing for innovative ways of preparing different classes of quantum states.

For example, for distributions that are efficiently computable classically there also exists efficient methods for state preparation \cite{rattew2022preparingarbitrarycontinuousfunctions, guseynov2025efficientexplicitcircuitquantum, lemieux2024quantumsamplingalgorithmsquantum}. While the CX gate count grows exponentially for exact state preparation of unstructured data \cite{grover2002creatingsuperpositionscorrespondefficiently, Mottonen2005Transformation, Shende2006Synthesis, Iten2016Quantum}, the circuit depth can be reduced with ancillary qubits \cite{Araujo2021divide, Zhang2022Quantum, Rattew2021efficient}.
The CX gate count can be lowered if the state is only required to be prepared approximately \cite{Maxwell2024Drastic, Zoufal2019, MarinSanchez2023Quantum, Benedetti_2019, Zylberman2024Efficient}. Advanced state preparation methods are also available for sparse \cite{Gleinig2022Efficient, Malvetti2021quantumcircuits, li2025nearlyoptimalcircuitsize, gaidai2025decompositionsparseamplitudepermutation}, Dicke \cite{yuan2025depthefficientquantumcircuitsynthesis, Aktar2022divide, Bartschi2019Deterministic}, uniform \cite{Mozafari2020Automatic, Mozafari2021Efficient, Mozafari2022Efficient} and low Schmidt rank states \cite{Araujo2024Low}. Some works have focused on preparing the state in some transformed basis \cite{Moosa_2024, Jobst2024efficientmps, warner2025quantumstatepreparationnested, Endo2020Quantum}.

\begin{figure*}
\centering
\begin{tikzpicture}
\begin{yquant}
qubit q[6];  
box {$U_{1}$} (q[0], q[1]);
box {$U_{2}$} (q[1], q[2]);
box {$U_{3}$} (q[2], q[3]);
box {$U_{4}$} (q[3], q[4]);
box {$U_{5}$} (q[4], q[5]);
box {$U_{6}$} (q[5]);
barrier q;
cnot q[1] | q[2];
cnot q[3] | q[4];
box {$R_y(\theta_{20})$} q[1];
box {$R_y(\theta_{19})$} q[2];
box {$R_y(\theta_{18})$} q[3];
box {$R_y(\theta_{17})$} q[4];
cnot q[0] | q[1];
cnot q[2] | q[3];
cnot q[4] | q[5];
box {$R_y(\theta_{16})$} q[0];
box {$R_y(\theta_{15})$} q[1];
box {$R_y(\theta_{14})$} q[2];
box {$R_y(\theta_{13})$} q[3];
box {$R_y(\theta_{12})$} q[4];
box {$R_y(\theta_{11})$} q[5];
cnot q[1] | q[2];
cnot q[3] | q[4];
box {$R_y(\theta_{10})$} q[1];
box {$R_y(\theta_{9})$} q[2];
box {$R_y(\theta_{8})$} q[3];
box {$R_y(\theta_{7})$} q[4];
cnot q[0] | q[1];
cnot q[2] | q[3];
cnot q[4] | q[5];
box {$R_y(\theta_{6})$} q[0];
box {$R_y(\theta_{5})$} q[1];
box {$R_y(\theta_{4})$} q[2];
box {$R_y(\theta_{3})$} q[3];
box {$R_y(\theta_{2})$} q[4];
box {$R_y(\theta_{1})$} q[5];
\end{yquant}
\end{tikzpicture}
\caption{A single truncated MPS disentangling layer followed by a PQC that minimizes entanglement entropy.}
\label{fig:ee+mps}
\end{figure*}

Matrix product states (MPS) provide a promising candidate to approximate quantum state preparation \cite{Ran2020Encoding, Malz2024Preparation, Rudolph_2024, BenDov2024Approximate, Melnikov_2023, Smith2024Cosntant, Schollwock2011TheDensity, Vidal2003Efficient, Fomichev2024Initial, Holmes2020Efficient, Cirac2021Matrix, Zhou2021Automatically, GonzalezConde2024efficientquantum, Gundlapalli2022Deterministic}. However, MPS representations of quantum states are typically truncated to reach reasonable scaling in terms of gate count. For arbitrary states the truncation error can become significant. MPS are also used, for example, in the context of many-body physics to simulate ground states of Hamiltonians \cite{White1992Density, Vidal2008Class, Evenbly2014, Fishman2015Compression, Hauschild2018Finding}.

Parametrized quantum circuits \cite{Schuld2019Quantum, Havlicek2019Supervised, Zoufal2019, Benedetti_2019} (PQCs) can also be used for approximate state preparation. A PQC consists of alternating layers of CX and single qubit parametrized rotation gates. The parameters can be trained to minimize distance to a target state by computing gradients \cite{luchnikov2021riemannian}, or by parameter-shift rules \cite{Mitarai2018Quantum, Schuld2019Evaluating}. The limitation with PQCs is the difficulty in training them as parametrized circuits suffer from the barren plateau problem \cite{Larocca2025}, and therefore they may be infeasible to train in practice.

\begin{figure}
    \centering
    \includegraphics[width=1\linewidth]{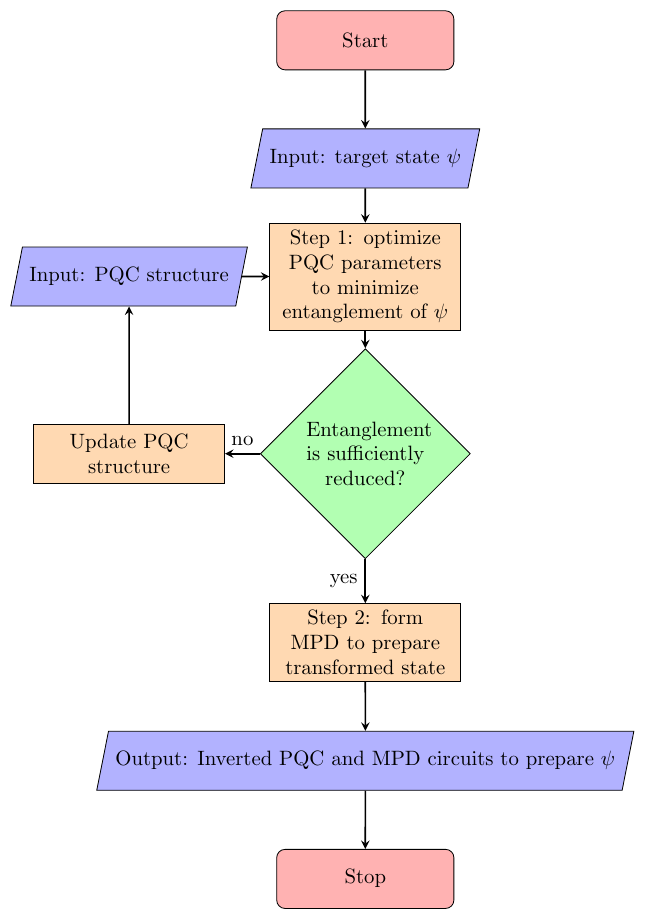}
    \caption{Two-step VDSP method for QSP. MPD stands for matrix product disentangler.}
    \label{fig:flow}
\end{figure}

We present an innovative two-step method for QSP. Our method is showcased in Fig. \ref{fig:flow}. Note that our method only requires connectivity between adjacent qubits which makes the method suitable for NISQ devices. In the first step, we train a PQC with the aim of minimizing the entanglement entropy of a quantum state $\ket{\psi}$. Thus, the PQC transforms the initial state $\ket{\psi}$ as $\ket{\psi} \mapsto \ket{\psi'}$, with $\ket{\psi'}$ having a lower entanglement entropy. In the second step, we use a shallow MPS disentangling circuit to prepare $\ket{\psi'}$. In the paper, we used a single MPS layer of bond dimension $\chi = 2$ as defined in Section~\ref{sec:mps} although we note that the approach is flexible and multiple MPD layers and of larger bond dimensions can also be used. These two steps are combined by inverting the PQC and appending it to the MPS disentangling circuit, as shown in Figure~\ref{fig:ee+mps}. We will call this method the \textit{variational disentangling state preparation} (VDSP) method. This two-step process allows for significantly increased accuracy in QSP compared to PQC or MPS circuits alone while using a comparable number of CX gates. Furthermore, in the Discussion section we note some arguments on the scalability of training the PQC. The usefulness and scalability of the presented method is demonstrated through various benchmarks.

The rest of the paper is organized as follows. In the next two sections we give a brief overview of matrix product states and parametrized circuits for QSP. We then introduce our novel two-step method and present accuracy guarantees based on entanglement entropy. The performance and accuracy of our method is demonstrated through selected benchmarks. Finally, we end with a brief discussion of possible future directions for further research. Additional details and benchmarks can be found in Appendix~\ref{appendix:benchmarks}. The Results section contains benchmarks for 2D normal distribution and 2D Ricker wavelet states. The full benchmark data can be found in Appendix~\ref{appendix:benchmarks}.

The choice of 2D normal distribution and Ricker wavelet states is driven by the desire to model smooth distributions that appear in fluid dynamic simulations \cite{bastidazamora2026quantumalgorithmlatticeboltzmann} as well as in particular the advection diffusion problem, where an initial gaussian distribution often evolves over time into a shape resembling the Ricker wavelet.

\section{Matrix product states}\label{sec:mps}

MPS provide a useful tool for QSP. First, an MPS representation of a quantum state is found by iteratively applying the singular value decomposition: 
\begin{align}\label{eq:MPS}
    \ket{\psi} = \sum_{i_1 \dots i_N \in \{0, 1\}^N} A^1_{i_1, a_1} A^2_{i_2, a_1 a_2} \dots A^N_{i_N, a_{N-1}} \ket{i_1 i_2 \dots i_N},
\end{align} where $A^1$ and $A^N$ are rank-2 tensors, $A^i$ a rank-3 tensor for $1<i<N$. The physical indices $i$ range over $\{0,1\}$ for qubits, while the virtual indices $a_i \in \{1, 2, \dots , \chi_i \}$ are summed over in neighboring tensors. The term ``bond'' is used to refer to any virtual index of the MPS, or a property relating to that virtual index. The number $\chi_i$ is called the bond dimension of site $i$, and the maximum bond dimension is denoted with $\chi$.  For arbitrary states $\chi$ grows exponentially in the number of qubits.

The reason why MPS are useful for QSP is that the representation \eqref{eq:MPS} can be directly transformed into a quantum circuit \cite{Ran2020Encoding} that prepares the state. However, for site $i$, it takes a unitary gate $U_i$ acting on $\ceil{\log_2 \dim(a_i)} + 1$ qubits to realize the circuit. By truncating the individual bond dimensions it is possible to find a quantum circuit that approximates the desired state while being efficient to implement as a quantum circuit. A matrix product disentangler (MPD) is built by truncating the bond dimensions of the representation \eqref{eq:MPS} to a suitably low number, typically to two so that the circuit can be implemented via two-qubit gates. Thus we obtain $\ket{\psi^1} = U^1_{\chi = 2} \ket{0}$, where $U^1_{\chi = 2}$ represents the truncated MPS. We can then calculate $\ket{\psi_2 '} = (U^1_{\chi = 2})^\dagger \ket{\psi}$ to iteratively disentangle the initial state $\ket{\psi}$ and build layers of MPDs to fully disentangle $\ket{\psi}$. This process can be repeated arbitrarily many times, until $\ket{\psi} \approx U^m_{\chi = 2} \dots U^2_{\chi = 2}U^1_{\chi = 2} \ket{0}$ sufficiently well approximates $\ket{\psi}$.

Unfortunately, the MPDs do not in practice converge to a good quality approximation of the target state unless the target state has low entanglement across the bonds. In the literature MPS are usually used for smooth distributions \cite{Iaconis2024}, and for these states the method works well. There also exist various ways to optimize MPS circuits \cite{Rudolph_2024, Smith2024Cosntant, bohun2025entanglementscalingmatrixproduct}, but we will use the basic circuits introduced in \cite{Ran2020Encoding} to demonstrate our method.

We will now introduce parametrized quantum circuits before analyzing the entanglement of a quantum state $\ket{\psi}$ across the bonds and show how we can improve the convergence of the approximation via our two-step process.

\section{Parametrized quantum circuits}

Parametrized quantum circuits (PQCs) are widely used in quantum machine learning \cite{Schuld2019Quantum, Havlicek2019Supervised, Zoufal2019, Benedetti_2019}. The goal is to train the parameters of a PQC to achieve good performance in some task. For a particular task, a particular form of parametrized circuit known as an ansatz is chosen to be suitable for the task. In this work, we focus on PQCs that require only a linear chain of nearest neighbor connectivity, known as a hardware efficient ansatz \cite{Kandala2017, Jose2022Error}. The exact way in which we organize a PQC into rotation and entangling layers is illustrated in Fig. \ref{fig:PQC}. Nota that each variational layer consists of two layers of CX gates; each layer contains one even and one odd layer of CX gates. We will refer to this layout of parametrized gates as the brick wall PQC. We let $U^N_m({\Vec{\theta}})$ denote the unitary representation of a PQC with $N$ qubits and $m$ parallel CX layers. The initial parameters are randomly sampled from a Gaussian distribution for optimal convergence \cite{NEURIPS2022_7611a3cb, kashif2023alleviatingbarrenplateausparameterized}.

Each CX layer is preceded by a layer of parametrized $R_y(\theta)$ gates, with $\Vec{\theta}$ being the vector of all the individual $(\theta)$ across all layers. For QSP of real-valued data the combination of CX and $R_y$ layers is sufficient. For state preparation tasks encoding data using complex amplitudes it is necessary to also include $R_x$ and $R_z$ gates to achieve arbitrary rotations.

The PQC $U^N_m(\Vec{\theta})$ consists of $m (N-1)$ CX gates and for each CX gate there are two $R_y$ gates. The PQC can be trained by calculating gradients of a loss function with respect to the variables $\Vec{\theta}$. For instance, the loss function can be chosen to minimize the Euclidean distance \eqref{eq:distance-loss} or to maximize the fidelity against a target state \eqref{eq:fidelity-loss}: \begin{align}
    \label{eq:distance-loss}
    L_{dist}(\Vec{\theta}, \psi) = \no{\psi - U^N_m(\Vec{\theta})\ket{0}}
\end{align}
\begin{align}
    \label{eq:fidelity-loss}
    L_{fid}(\Vec{\theta}, \psi) = 1 - \mo{\bra{\psi}U^N_m(\Vec{\theta})\ket{0}}^2
\end{align}
where $\no{\cdot}$ is the Euclidean norm. We will use $1-L_{dist}$ as our metric for \textit{accuracy}, while to Eq. \eqref{eq:fidelity-loss} we will refer to as \textit{infidelity}. The gradients with respect to the chosen loss function can be calculated using the quantum natural gradient \cite{luchnikov2021riemannian}, or by parameter-shift rules \cite{Mitarai2018Quantum, Schuld2019Evaluating}. We used \texttt{quimb} \cite{Gray2018} for training which allowed us to both train the ansatz and manipulate matrix product states and operators with a single tool. Ee used the loss function Eq. \eqref{eq:distance-loss} exclusively when training PQCs  directly for state preparation.

\begin{figure*}
\centering
\begin{tikzpicture}
\begin{yquant}
qubit q[4];  

box {$R_y(\theta_{16})$} q[0];
box {$R_y(\theta_{15})$} q[1];
box {$R_y(\theta_{14})$} q[2];
box {$R_y(\theta_{13})$} q[3];
cnot q[0] | q[1];
cnot q[2] | q[3];
box {$R_y(\theta_{12})$} q[1];
box {$R_y(\theta_{11})$} q[2];
cnot q[1] | q[2];
barrier q;
box {$R_y(\theta_{10})$} q[0];
box {$R_y(\theta_{9})$} q[1];
box {$R_y(\theta_{8})$} q[2];
box {$R_y(\theta_{7})$} q[3];
cnot q[0] | q[1]; 
cnot q[2] | q[3];
box {$R_y(\theta_{6})$} q[1];
box {$R_y(\theta_{5})$} q[2];
cnot q[1] | q[2];
barrier q;
box {$R_y(\theta_{4})$} q[0];
box {$R_y(\theta_{3})$} q[1];
box {$R_y(\theta_{2})$} q[2];
box {$R_y(\theta_{1})$} q[3];
\end{yquant}
\end{tikzpicture}
\caption{A brick wall PQC with 4 qubits and two layers.}
\label{fig:PQC}
\end{figure*}

\section{Minimizing entanglement entropy}

Next we will define the entanglement entropy so that we can use it as a loss function with a PQC. Our definition of entanglement entropy is closely related to that of von Neumann entropy. Recall Eq. \eqref{eq:MPS}. Instead of writing the state $\ket{\psi}$ in terms of the tensors directly, we can keep track of the singular values when we iteratively perform SVDs to form the MPS representation. Expressed in this way, Eg. \eqref{eq:MPS} becomes \begin{align}\label{eq:MPS-SVD}
    \ket{\psi} = \sum_{\{i\}} \Gamma^1_{i_1,a_1} \Lambda^1_{a_1} \Gamma^2_{i_2,a_1a2} \Lambda^2_{a_2} \dots \Lambda^{N-1}_{a_{N-1}} \Gamma^N_{i_n, a_{N-1}}\ket{i},
\end{align} where $\Lambda^i$ are the singular values as found by forming the expression \eqref{eq:MPS} and $\Gamma^i$ are ``bare'' tensors from the SVD decompositions. This form of the MPS can be easily transformed to left- or right orthogonal forms by absorbing the singular values to the bare tensors on either the left or right hand side \cite{Schollwock2011TheDensity}. The quantity \begin{align}
    S_k = -\sum_{i=1} (\Lambda^k_i)^2 \log_2 ((\Lambda^k_i)^2) 
\end{align} is the entanglement entropy at bond $k$ for the state $\ket{\psi}$. It expresses how entangled the state is when partitioned to two parts, one part to the left and one part to the right of the bond $k$. Observe that, if all but the first two terms of $S_k$ are zero, then the corresponding tensor can be perfectly transformed to a two-qubit gate. In order to find an efficient representation of $\ket{\psi}$ as a quantum circuit, we thus define the cumulative entanglement entropy of a state $\ket{\psi}$ as \begin{align}\label{eq:ee-loss}
    S(\ket{{\psi}}) = \sum_{k, i>2} (\Lambda^k_i)^2 \log_2 ((\Lambda^k_i)^2).
\end{align} For the rest of the paper we will use Eq. \eqref{eq:ee-loss} as the definition of entanglement entropy of a state and we report this value for all of the benchmarks in the Appendix \ref{appendix:benchmarks}. This is a positive quantity that reaches the value zero when the state can be perfectly expressed as a single MPD layer with $\chi = 2$. Minimizing this quantity thus directly optimizes how the state can be approximated via MPDs. However, it may be better to use a modified version of Eq. \eqref{eq:ee-loss} for training. We define the linear entanglement entropy \begin{align}\label{eq:ee-loss-linear}
    S(\ket{{\psi}}) = \sum_{k, i>2} \Lambda^k_i
\end{align} which should provide better gradients for the optimizer and use that as the loss function for all training runs. Figure~\ref{fig:ee+mps} shows how the quantum circuit looks like in practice. Notice that the order of the PQC and MPD layers are inverted in the final circuit, as the final circuit inverts the PQC and MPD layers.

As an example, let us demonstrate how the method works by preparing a 2D normal distribution: $\mathcal{N}(\Vec{\mu}, \Sigma)$, where $\Vec{\mu}$ defines the mean and $\Sigma$ is the covariance matrix. We use \begin{align}
    \Vec{\mu} = \begin{bmatrix}
    0.5 & 0.5
\end{bmatrix}, \quad \Sigma = \begin{bmatrix}
    0.1 & 0.01 \\ 0.01 & 0.1
\end{bmatrix}.
\end{align} For this example we will use 10 qubits, so we use 5 qubits to discretize the $x$ and $y$ dimensions, respectively. We use a binary decimal representation for each dimension. For instance, in the case of 3 qubits, the state $\ket{010}$ would correspond to the data point $0 \cdot 2^{-1} + 1 \cdot 2^{-2} + 0 \cdot 2^{-3} = \frac{1}{4}.$ In this way we can discretize the grid $[0,1) \times [0, 1)$ into evenly spaced points, with both dimensions being discretized into 32 points. Note that the end point $1$ is not included in the interval. We note that some works discretize intervals to also include the end point \cite{Iaconis2024}, which is a subtle difference.

For this example $\Vec{\mu}$ and $\Sigma$ were chosen to get a distribution roughly centered at the middle of the grid.  We trained a 2-layer PQC to minimize entanglement entropy, which reduced it from $0.0055$ to $0.0014$, which is roughly a 75\% reduction. We then used a single layer MPD with $\chi=2$ to prepare the transformed state. We achieved accuracy of $0.9931$, when the infidelity was $4.71 \times 10^{-5}$, with a circuit consisting of 40 CX gates. The full training procedure is illustrated in Appendix~\ref{appendix:benchmarks}.

Before reporting the benchmark results, let us first analyze the relationship between entanglement entropy and truncation error of MPS.

\subsection{Truncation error of MPS}

When truncating the bond dimension of an MPS to $k$, the truncation error is bounded by the leading singular values \cite{manabe2025statepreparationmultivariatenormal, Oseledets2011Tensor, Oseledets2010TTCross}:
\begin{align}\label{eq:mps-approx-error}
    \no{\psi - U^1_{\chi=k}\ket{0}} & \leq \sqrt{\sum_{i=1}^{n-1}\epsilon^2_i},\\
    \epsilon^2_i = 1 - \sum_{j=1}^{k} (\Lambda_j^{i})^2,
\end{align}
Thus, by minimizing cumulative entanglement entropy, we will reduce the truncation error of the corresponding MPS for the transformed state. We now prove that this is also equivalent to the final approximation error with the PQC included for the target state. Namely, suppose we want to prepare the target state $\ket{\psi}$, let $U^N_m(\Vec{\theta})$ be the PQC that minimizes entanglement entropy, and let $U_{\chi = 2}$ be an MPD unitary. Then, \begin{align}\label{eq:truncation-error}\begin{split}
    \no{\psi - U^N_m(\Vec{\theta}) U_{\chi=2} \ket{0}} &= \no{U^N_m(\Vec{\theta})\left(U^{N\dagger}_m(\Vec{\theta}) \psi -   U_{\chi=2} \ket{0} \right)} \\
    & = \no{U^{N\dagger}_m(\Vec{\theta}) \psi -  U_{\chi=2} \ket{0}}
    \end{split}
\end{align}
Eq. \eqref{eq:truncation-error} proves that the final accuracy of our method is lower bounded by the MPS approximation error of Eq. \eqref{eq:mps-approx-error} for the transformed state with minimized entanglement entropy. It is also possible to further improve accuracy by increasing the MPD layers, although additional layers typically improve accuracy only marginally.

\subsection{Lower bounds on number of variational layers}\label{sec:theory}

Before presenting benchmarks of the VDSP method we seek to find some theoretical explanation of how many layers we need to disentangle a state. Suppose the target state $\ket{\psi}$ is a MPS with bond dimension $\chi = D_\psi$. The goal of the VDSP method is to lower the truncation error of Eq. \eqref{eq:mps-approx-error}, that is, the method attempts to variationally find a unitary transformation $U$ such that $\ket{\psi'} = U \ket{\psi}$ has a low approximation error to a chosen dimension $k$. In this work we always truncate to dimension $k=2$. 

There are at least two mechanisms by which the objective of lowering the truncation error can be achieved. Proposition~\ref{prop:brickwall-layers} presents lower bound on variational layers based on a parameter counting argument for exact state preparation, and thus this lower bound represents the strongest bound on scaling in terms of the number of variational layers required. We then make a few observations on how the Schmidt rank and bond dimensions of a state can be changed by a two-qubit unitary gate. These observations lead to Proposition~\ref{prop:rank-reduction}, which represents a lower bound on variational layers based on the ability of a quantum gate to reduce the Schmidt rank of the state. Propositions \ref{prop:brickwall-layers} and \ref{prop:rank-reduction} thus work with different assumptions. Finally, Theorem~\ref{theorem:interference} explains the mechanism of entanglement reduction by direct calculation.

An arbitrary $N$-qubit pure state has $2^{N+1}-2$ real parameters, while a MPS with bond dimension $\chi = D_\psi$ and physical dimension 2 has roughly $2ND_{\psi}^2$ independent real parameters \cite{Schollwock2011TheDensity}. The VDSP method is looking for a transformation that transforms the target state approximately into a state with bond dimension two. We can argue that the VDSP could work by reducing the number of parameters in the target state from $2ND_{\psi}^2$ to $2N2^2$. Thus it is required to reduce the number of parameters in the state roughly by $2N(D_{\psi}^2-4)$. Each CX gate can introduce 4 parameters into a state, so it must be also true that a single CX gate can remove at most 4 parameters from a state. Thus the number of CX gates to reduce the MPS bond dimension from $D_{\psi}$ to 2 must be at least $N(D_{\psi}^2-4)/2$. Each variational layer in the brick wall PQC has $N-1$ CX gates, thus we can approximate the required number of layers by $(D_{\psi}^2-4)/2$. We state these observations in the following proposition.

\begin{proposition}\label{prop:brickwall-layers}
    A unitary transformation that maps a MPS with bond dimension $\chi = D$ and $2ND^2$ independent real parameters into a MPS with $\chi = 2$ requires at least $N(D^2-4)/2$ CX gates. A brick wall PQC that implements this transformation requires at least $(D^2-4)/2$ layers.
\end{proposition}

Proposition~\ref{prop:brickwall-layers} represents a lower bound on the number of CX gates based on a parameter counting argument and thus represents the strongest lower bound on variational layers.  Note that this bound assumes we can perfectly train the PQC without getting stuck to local minima, and that the brick wall PQC is expressive enough to be able to implement the required transformation. We will not attempt to prove either of these assumptions, and instead take this bound as a rough lower bound on the resources required to achieve exact state preparation.

There is another possible way the VDSP could be able to reduce truncation error. Instead of counting parameters in the state and arguing how they may be reduced, we may directly observe how a two-qubit gate acts on a MPS. In this section we will often use the notation $U\psi$ when $U$ is a two-qubit unitary gate and $\psi$ is an $n$-qubit state. The target qubits on which $U$ acts are always clear from context.

\begin{lemma}\label{lemma:gate-effect}
    Let $\psi$ be an $n$-qubit quantum state and $U$ a unitary gate that acts on qubits $j, j+1$. Denote by $\{ \lambda_i^{(k)} (\psi) \}_i$ the Schmidt values of the Schmidt decomposition \begin{align}
        \psi = \sum_{i=1}^{\chi_k} \lambda^{k}_i \psi^L_i \otimes \psi^R_i,
    \end{align}where $\psi^L_i \in \mathcal{H}^L$, $\psi^R_i \in \mathcal{H}^R$ and $\mathcal{H}^L$, $\mathcal{H}^R$ are the Hilbert spaces of qubits $\{ 1, \dots , k \}$ and $\{ k+1, \dots , n \}$, respectively. The Schmidt rank (according to bipartition into $\mathcal{H}^L$ and $\mathcal{H}^R$) of the state is $\chi_k$. For any value $k \neq j$, the Schmidt values satisfy $\{ \lambda_i^{(k)} (\psi) \}_i = \{ \lambda_i^{(k)} (U\psi) \}_i$.
\end{lemma}
\begin{proof}
    If we form the Schmidt decomposition at any cut $k < j$, we have \begin{align}
        U\psi = \sum_{i=1}^{\chi_k} \lambda^{k}_i \psi^L_i \otimes U\psi^R_i,
    \end{align}so the unitary $U$ acts only on the space $\mathcal{H}^R$. A local unitary only rotates the Schmidt vectors and leaves the Schmidt values unchanged. The same argument holds for $k > j$ but the unitary $U$ only acts on $\mathcal{H}^L$.
\end{proof}

Note that the Schmidt rank and Schmidt values coincide with the bond dimension and singular values of the state in (canonical) MPS form of Eq.~\eqref{eq:MPS-SVD}. Thus we have shown by Lemma \ref{lemma:gate-effect} that a two-qubit gate acts on an isolated bond in a MPS. A natural question then arises: how much can a single two-qubit gate increase or decrease the entanglement in a MPS? A tight bound on this is given by the following structural observation.

\begin{proposition}\label{prop:structure}
    Let $\psi$ be an $n$-qubit quantum state. Let \begin{align}\label{eq:prop-schmidt-cut}
        \psi = \sum_{i=1}^{\chi_k} \lambda^{k}_i \psi^L_i \otimes \psi^R_i,
    \end{align} be its Schmidt decomposition at cut $k$. The Schmidt rank $\chi_k$ is bounded by Schmidt ranks at adjacent cuts by \begin{align}
       \max \left( \ceil{\frac{\chi_{k-1}}{2}}, \ceil{\frac{\chi_{k+1}}{2}} \right) \leq \chi_k \leq \min \left( 2 \chi_{k-1}, 2 \chi_{k+1} \right).
    \end{align}
\end{proposition}
\begin{proof}
    We can write each vector $\psi^R_i$ as \begin{align}
        \psi^R_i = \sum_{a=1}^2 \sigma_{i,a}\varphi_a \otimes \xi_{i,a},
    \end{align} where $\{ \varphi_i \}$ forms a basis for the qubit $k+1$. Repartitioning according to the cut $k+1$ gives \begin{align}
        \psi = \sum_{i=1}^{\chi_k}\sum_{a=1}^2 \lambda^{k}_i \sigma_{i,a} \left(\psi^L_i \otimes \varphi_a \right) \otimes \xi_{i,a},
    \end{align} which shows that $\chi_{k+1} \leq 2\chi_k$. The same argument in the opposite direction gives $\chi_{k} \leq 2\chi_{k+1}$, or $\chi_{k+1} \geq \ceil{\frac{\chi_{k}}{2}}$. The proof is completed by repartitioning the Schmidt decomposition \eqref{eq:prop-schmidt-cut} according to cut $k-1$.
\end{proof}

The structural result of Proposition~\ref{prop:structure} always holds, but it does not tell much about the effect of a gate on the Schmidt rank of a quantum state. It gives the physical limits of what the Schmidt rank can be. We can combine the result with the action of a gate to obtain a more informative bound. The operator Schmidt decomposition \cite{Nielsen2003Quantum, Bengtsson_Zyczkowski_2006, MULLERHERMES2018174} is used in the following proposition.

\begin{proposition}\label{prop:operator-schmidt-rank}
    Let $\psi$ be an $n$-qubit quantum state with Schmidt decomposition \begin{align}
        \psi = \sum_{i=1}^{\chi_k} \lambda^{k}_i \psi^L_i \otimes \psi^R_i
    \end{align} at cut $k$. Suppose $U$ is a two-qubit unitary gate with operator Schmidt rank $r$, that acts on qubits $k$ and $k+1$. Then the Schmidt rank $\chi'_k$ of the state $U\psi$ at cut $k$ satisfies \begin{align}
       \chi'_k  &\geq \max \left( \ceil{\frac{\chi_{k-1}}{2}}, \ceil{\frac{\chi_{k+1}}{2}}, \ceil{\frac{\chi_{k}}{r}} \right) \\
        \chi'_k &\leq \min \left( 2 \chi_{k-1}, 2 \chi_{k+1}, r\chi_k \right).
    \end{align}
\end{proposition}
\begin{proof}
    Let \begin{align}
        \psi = \sum_{i=1}^{\chi_k} \lambda^{k}_i \psi^L_i \otimes \psi^R_i,
    \end{align} be the Schmidt decomposition of $\psi$ at cut $k$ and \begin{align}
        U = \sum_{j=1}^r \sigma_j A_j \otimes B_j,
    \end{align} where $\{A_i\}$ and $\{B_i\}$ are orthonormal sets of operators and $\{\sigma_j\}$ are the operator Schmidt values. Then \begin{align}
        U\psi = \sum_{i=1}^{\chi_k}\sum_{j=1}^r \lambda_i \sigma_j \left(A_j \psi^L_i \right) \otimes \left(B_j \psi^R_i \right).
    \end{align} Both sets of vectors $\{A_j \psi^L_i\}$ and $\{B_j \psi^R_i\}$ span at most a $r\chi_k$ dimensional space, so it follows that $\chi'_k \leq r \chi_k$. Applying the operator $U^\dagger$ to the Schmidt rank of $U\psi$ together with Proposition~\ref{prop:structure} completes the proof.
\end{proof}

Based on the structural limitations on MPS, we can now state the following limitation on the action of a single two-qubit gate on MPS.

\begin{corollary}\label{corollary:brick-wall}
    Let $\psi$ be a $n$-qubit quantum state with a fixed bond dimension $D$ at sites $k-1$, $k$ and $k+1$. After applying a two-qubit gate with operator Schmidt rank $r$ to qubits $k$ and $k+1$, the new bond dimension $D'_k$ at site $k$ satisfies $\ceil{\frac{D}{r}} \leq D'_k \leq rD $. To increase or decrease $D'_k$ further, a two-qubit gate has to be applied to both qubit pairs $\{ k-1, k \}$ and $\{ k+1,k+2 \}$.
\end{corollary}

Corollary~\ref{corollary:brick-wall} is a direct consequence of Lemma~\ref{lemma:gate-effect} and Proposition~\ref{prop:operator-schmidt-rank}. In order to effectively decrease the bond dimension of a MPS, and perhaps more crucially in order to decrease the truncation error to a state of lower bond dimension, it is necessary to apply two-qubit gates in a systematic manner. A single CX gate has operator Schmidt rank 2 and acts on an isolated bond. Thus it can only increase or decrease the bond dimension, or Schmidt rank, by a factor of two. For subsequent reduction in Schmidt rank, it is necessary to act on both adjacent sites. This strongly suggests that a brick wall PQC has the best possible arrangement of gates with respect to circuit depth when the qubits are limited by linear connectivity. Based on the reduction in Schmidt rank, we can state the following lower bound on the required number of layers in a brick wall PQC to reduce a MPS into a state with bond dimension two.

\begin{proposition}\label{prop:rank-reduction}
    Let $\psi$ be an $n$-qubit state with bond dimension $D$. A transformation that maps $\psi$  to a state with bond dimension 2 must consists of at least $\lfloor \log_2 \frac{D}{2} \rfloor$ CX gates acting on all sites with bond dimension $D$. A brick wall PQC that implements this transformation must consist of at least $\lfloor \log_2 \frac{D}{2} \rfloor$ layers.
\end{proposition}
\begin{proof}
    Proposition \ref{prop:operator-schmidt-rank} lays the foundation of how much a single CX gate can do. Namely, as a CX gate has operator Schmidt rank two, a single CX gate can only change the Schmidt rank at a specific cut by a factor of two. The brick wall PQC has a single CX gate per cut from which the lower bound follows.
\end{proof}

We have presented two lower bounds on the number of required variational layers in the VDSP method. Proposition~\ref{prop:brickwall-layers} is based on a parameter counting argument, while Proposition~\ref{prop:rank-reduction} is based on rank reduction and observing a single CX gate has limited ability to reduce the Schmidt rank. However, neither of these lower bounds explain the actual mechanics by which a quantum gate is able to reduce entanglement entropy of a state.

We can actually calculate the Schmidt value spectrum of an updated state directly to explain the exact entanglement reducing mechanism explicitly.


\begin{theorem}\label{theorem:interference}
    Let $\psi$ be an $n$-qubit quantum state with Schmidt decomposition \begin{align}
        \psi = \sum_{i=1}^{\chi_k} \lambda_i \psi^L_i \otimes \psi^R_i
    \end{align} at cut $k$ and let $U$ be a two-qubit unitary operator acting on qubits $k$ and $k+1$ with operator Schmidt decomposition \begin{align}
        U = \sum_{\alpha=1}^r \sigma_\alpha A_\alpha \otimes B_\alpha.
    \end{align}
    After applying $U$ to the state $\psi$, the Schmidt values of the updated state are the singular values of the bipartition matrix \begin{align}\label{eq:matrix-equation}
        M = P D Q^T,
    \end{align} where $D = \mathrm{diag}(\Vec{\lambda} \otimes \Vec{\sigma})$, $\Vec{\lambda} \otimes \Vec{\sigma}$ is the Kronecker product of vectors $\Vec{\lambda} = \begin{bmatrix}
        \lambda_1 & \dots & \lambda_{\chi_k}
    \end{bmatrix}$ and $\Vec{\sigma} = \begin{bmatrix}
        \sigma_1 & \dots & \sigma_r
    \end{bmatrix}$, and $P$ is a $2^k \times r\chi_k$ matrix with columns $\{(\id_{2^{k-1}} \otimes A_\alpha)\psi^L_i\}_{i,\alpha}$ and $Q$ is a $2^{n-k} \times r\chi_k$ matrix with columns $\{(B_\alpha \otimes \id_{2^{n-k-1}})\psi^R_i\}_{i, \alpha}$. 
\end{theorem}

\begin{proof}
    Since $U$ is a two-qubit gate, let us denote by $\Tilde{U} = \id_{2^{k-1}} \otimes U \otimes \id_{2^{n-k-1}}$ the operator with correct dimensions. Then \begin{align}
        \Tilde{U}\psi = \sum_{i=1}^{\chi_k} \sum_{\alpha=1}^r \lambda_i \sigma_\alpha \Tilde{A}_\alpha \psi^L_i \otimes \Tilde{B}_\alpha \psi^R_i,
    \end{align} where $\Tilde{A}_\alpha = \id_{2^{k-1}} \otimes A_\alpha$ and $\Tilde{B}_\alpha = B_\alpha \otimes \id_{2^{n-k-1}}$. A bipartition matrix $M$ of a generic state $\varphi$ is defined following $\varphi = \sum_{\mu,\nu}M_{\mu \nu}\ket{\mu} \ket{\nu}$, where $\{\ket{\mu}\}$ and $\{\ket{\nu}\}$ are orthonormal bases of the left and right partitions, and the singular values of $M$ coincide with the Schmidt values of $\varphi$. We can write \begin{align}
        \Tilde{U}\psi & = \sum_{i=1}^{\chi_k}\sum_{\alpha=1}^r\sum_{\mu=1}^{2^k}\sum_{\nu=1}^{2^{n-k}}\lambda_i \sigma_\alpha \ip{\mu}{\Tilde{A}_\alpha \psi^L_i} \ket{\mu} \otimes \ip{\nu}{\Tilde{B}_\alpha \psi^R_i} \ket{\nu} \\
        & = \sum_{\mu=1}^{2^k}\sum_{\nu=1}^{2^{n-k}}\sum_{i=1}^{\chi_k}\sum_{\alpha=1}^r \lambda_i \sigma_\alpha \ip{\mu}{\Tilde{A}_\alpha \psi^L_i} \ip{\nu}{\Tilde{B}_\alpha \psi^R_i} \ket{\mu}\otimes \ket{\nu}
    \end{align}
    so the bipartition matrix of $\Tilde{U}\psi$ can be written as \begin{align}
        M = P D Q^T,
    \end{align} where $P$ is a $2^k \times r\chi_k$ matrix with columns $\{\Tilde{A}_\alpha\psi^L_i\}$, $Q$ is a $2^{n-k} \times r\chi_k$ matrix with columns $\{\Tilde{B}_\alpha \psi^R_i\}$ and $D = \mathrm{diag}(\Vec{\lambda} \otimes \Vec{\sigma})$ is a diagonal matrix containing the Schmidt values of $\psi$ and the operator Schmidt values of $U$. 
\end{proof}

In the following corollary we highlight that the difficulty of calculating the updated Schmidt values only depends on the Schmidt rank of the quantum state and the operator Schmidt rank of the unitary operator.
\begin{corollary}\label{corollary:eigenvalue-problem}
    The singular values of the matrix $M$ in Eq. \eqref{eq:matrix-equation} can be calculated as an eigenvalue problem of size $r\chi_k \times r\chi_k$.
\end{corollary}
\begin{proof}
    We can calculate the squared singular values of $M$ as the eigenvalues of the matrix $M^\dagger M$: \begin{align}
        M^\dagger M &= (PDQ^T)^\dagger (PDQ^T) = \bar{Q}DP^\dagger P D Q^T \\
        & = \bar{Q}D G_LDQ^T,
    \end{align}where we define the left Gram matrix $G_L = P^\dagger P$. Note that the nonzero eigenvalues of $\bar{Q}D G_LDQ^T$ and $Q^T\bar{Q}D G_LD$ coincide. Furthermore, \begin{align}
        Q^T\bar{Q} = Q^T(Q^\dagger)^T = (Q^\dagger Q)^T = G_R^T,
    \end{align} where $G_R = Q^\dagger Q$ is the right Gram matrix. We then obtain the squared singular values of $M$ as the eigenvalues of the matrix \begin{align}\label{eq:schmidt-values-eigenvalue-problem}
        K = DG_LDG_R^T
    \end{align} which is a matrix of size $r\chi_k \times r \chi_k$.
\end{proof}

\begin{remark}
    The Gram matrices $G_L$ and $G_R$ have a very specific from. They can be written as $(G_L)_{(i,\alpha), (j, \beta)} = \langle \psi^L_i | \id \otimes A_\alpha^\dagger A_\beta | \psi^L_j \rangle$ and $(G_R)_{(i, \alpha), (j, \beta)} = \langle \psi^R_i |B_\alpha^\dagger B_\beta \otimes \id| \psi^R_j \rangle$. 
    Suppose that $G_L, G_R = \id$. Then if follows
    from Equation~\eqref{eq:schmidt-values-eigenvalue-problem} that the new Schmidt values are redistributed according to the Kronecker product $\Vec{\lambda} \otimes \Vec{\sigma}$. The Schmidt values are always positive, so this would mean entanglement cannot be decreased by such operation. In order to reduce entanglement it is required to carefully choose a unitary operator such that the operator Schmidt decomposition elements and state Schmidt decomposition elements interplay in such a way that the updated Schmidt values are concentrated on as few elements as possible.
\end{remark}

Theorem~\ref{theorem:interference} explains the mechanism by which a unitary gate can increase or reduce entanglement entropy in a quantum state. The amount of ``Schmidt mass'' a unitary gate can move is determined by Equation~\eqref{eq:matrix-equation}, or by Equation~\eqref{eq:schmidt-values-eigenvalue-problem} in more compact form. The Gram matrices $G_L$ and $G_R$ contain terms involving the operator Schmidt decomposition basis and the state Schmidt decomposition vectors. There is clearly a kind of interference going on between these elements -- constructive interference can increase entanglement and destructive interference reduces it. Remarkably, the question how much entanglement is increased or decreased by a specific two-qubit unitary gate is reduced to calculating the eigenvalues of a matrix of size $\chi r \times\chi r$ for a state with Schmidt rank $\chi$ and unitary operator with operator Schmidt rank $r$. This would potentially allow to optimize the disentangling gates quite efficiently in a greedy algorithm. However, it is unclear how effective a greedy algorithm would be at reducing entanglement between multiple qubits. While Equation~\eqref{eq:matrix-equation} explains the mechanism of entanglement reduction, a global optimizer might move the ``Schmidt mass'' of a quantum state with operations that are not strictly non-decreasing its entanglement. We study how the Schmidt mass is moved by the VDSP method in Section \ref{section:benchmark-analysis}. While Theorem~\ref{theorem:interference} alone might not explain exactly how the VDSP method works, it might be possible to design a very efficient greedy algorithm based on it. We leave this direction for future work.


\section{Results}\label{section:results}

In the results, we first present benchmark results using a noiseless statevector simulator to compare the methods with up to 20 qubits. To investigate the impact of noise on the different state preparation methods, we then present a noise analysis scaling the impact of noise. Finally, we report hardware results on a superconducting quantum computer.

Today, there exist many different software development kits for quantum computing, such as \texttt{qiskit}~\cite{qiskit2024} and \texttt{PennyLane}~\cite{pennylane2022}. In this paper we used \texttt{qiskit} for transpiling quantum circuits and \texttt{quimb}~\cite{Gray2018} for training the PQCs.

\subsection{Benchmark results}

We now report the accuracy of the VDSP method for the 2D normal distribution and the 2D Ricker wavelet states. The full benchmark data can be found in Appendix~\ref{appendix:benchmarks}. The 2D normal distribution is defined in the standard way as $\mathcal{N}(\Vec{\mu}, \Sigma)$, where $\Vec{\mu}$ defines the mean and $\Sigma$ is the covariance matrix. We use \begin{align}
    \Vec{\mu} = \begin{bmatrix}
    0.5 & 0.5
\end{bmatrix}, \quad \Sigma = \begin{bmatrix}
    0.1 & 0.01 \\ 0.01 & 0.1
\end{bmatrix}
\end{align} and simply flatten the 2D array and treat it as a 1-dimensional vector, although we acknowledge more sophisticated treatment of multi-dimensional arrays is possible \cite{Rosenkranz2025quantumstate, manabe2025statepreparationmultivariatenormal, GarciaRipoll2021quantuminspired}. However, as we will see, our approach to minimize entanglement entropy is also highly effective for the flattened 2D distributions. The 2D Ricker wavelet is defined as
\begin{align}
    \psi(x, y) = \frac{1}{\pi \sigma^4} \left( 1 - \frac{1}{2} \left( \frac{x^2 + y^2}{\sigma^2} \right) \right)e^{- \frac{x^2 + y^2}{2 \sigma^2}},
\end{align} which we discretized on $[-1, 1) \times [-1, 1)$ and used $\sigma=0.15$.
We train the PQCs with 1--5 layers with the linear entanglement entropy loss defined in Eq. \eqref{eq:ee-loss-linear}. More layers will generally speaking allow for greater fidelity, but for NISQ devices the circuits are preferably kept shallow. The range 1--5 allows us to analyze how fidelity is reduced while keeping the training manageable. 

In Appendix \ref{appendix:mpo} we provide analysis of how the bond dimension of the matrix product operator \cite{Schollwock2011TheDensity} representation of a PQC scales with the number of layers. We found that a linear number of layers (equal to the number of qubits) seems to achieve full bond dimension. However, a linear number of layers can be challenging to train. It is better to start with a shallow circuit and increase the number of layers until sufficient accuracy is reached. This is the approach we used. In general, the more qubits in the circuit, the more layers are required to disentangle the state effectively. The VDSP method is compared to exact state preparation \cite{Plesch2011Quantum} in terms of gates, and to a parametrized circuit of the same size, and to a MPD circuit \cite{Ran2020Encoding} with the same number disentangling layers.

\begin{table}[ht]
\centering
\begin{tabular}{l r r r r r}
\toprule
Method & $n_q$ & Accuracy & Infidelity & Depth & CX \\
\midrule
VDSP & 6 & 0.9987 & $1.81 \times 10^{-6}$ & 67 & 23 \\
PQC & 6 & 0.9770 & $5.28 \times 10^{-4}$ & 26 & 10 \\
MPD & 6 & 0.9915 & $7.20 \times 10^{-5}$ & 73 & 27 \\
Exact & 6 & 1.0000 & $0$ & 122 & 45 \\
\midrule
VDSP & 8 & 0.9964 & $1.30 \times 10^{-5}$ & 83 & 31 \\
PQC & 8 & 0.9406 & $3.53 \times 10^{-3}$ & 26 & 14 \\
MPD & 8 & 0.9845 & $2.40 \times 10^{-4}$ & 88 & 38 \\
Exact & 8 & 1.0000 & $0$ & 409 & 200 \\
\midrule
VDSP & 10 & 0.9931 & $4.71 \times 10^{-5}$ & 101 & 40 \\
PQC & 10 & 0.9373 & $3.93 \times 10^{-3}$ & 26 & 18 \\
MPD & 10 & 0.9848 & $2.31 \times 10^{-4}$ & 102 & 48 \\
Exact & 10 & 1.0000 & $0$ & 1696 & 876 \\
\midrule
VDSP & 12 & 0.9915 & $7.20 \times 10^{-5}$ & 118 & 50 \\
PQC & 12 & 0.9330 & $4.49 \times 10^{-3}$ & 26 & 22 \\
MPD & 12 & 0.9845 & $2.41 \times 10^{-4}$ & 122 & 60 \\
Exact & 12 & 1.0000 & $0$ & 6695 & 3613 \\
\midrule
VDSP & 14 & 0.9905 & $9.02 \times 10^{-5}$ & 135 & 61 \\
PQC & 14 & 0.9322 & $4.59 \times 10^{-3}$ & 26 & 26 \\
MPD & 14 & 0.9861 & $1.94 \times 10^{-4}$ & 138 & 64 \\
Exact & 14 & 1.0000 & $0$ & 27051 & 14761 \\
\midrule
VDSP & 16 & 0.9924 & $5.79 \times 10^{-5}$ & 154 & 67 \\
PQC & 16 & 0.9317 & $4.66 \times 10^{-3}$ & 26 & 30 \\
MPD & 16 & 0.9871 & $1.68 \times 10^{-4}$ & 150 & 73 \\
Exact & 16 & 1.0000 & $0$ & 108149 & 59517 \\
\midrule
VDSP & 18 & 0.9925 & $5.58 \times 10^{-5}$ & 171 & 82 \\
PQC & 18 & 0.9312 & $4.73 \times 10^{-3}$ & 26 & 34 \\
MPD & 18 & 0.9839 & $2.58 \times 10^{-4}$ & 169 & 87 \\
Exact & 18 & 1.0000 & $0$ & 433835 & 239304 \\
\midrule
VDSP & 20 & 0.9922 & $6.01 \times 10^{-5}$ & 184 & 85 \\
PQC & 20 & 0.9064 & $8.74 \times 10^{-3}$ & 26 & 38 \\
MPD & 20 & 0.9853 & $2.17 \times 10^{-4}$ & 186 & 97 \\
Exact & 20 & 1.0000 & $0$ & 1735314 & 959010 \\
\bottomrule
\end{tabular}
\caption{State preparation benchmark for the 2D normal distribution ($L_\mathrm{var}=2$, $L_\mathrm{MPS}=2$.)}
\label{tab:2d-normal-main}
\end{table}

\begin{table}[ht]
\centering
\begin{tabular}{l r r r r r}
\toprule
Method & $n_q$ & Accuracy & Infidelity & Depth & CX \\
\midrule
VDSP & 6 & 0.9994 & $4.21 \times 10^{-7}$ & 99 & 37 \\
PQC & 6 & 0.9998 & $6.08 \times 10^{-8}$ & 56 & 25 \\
MPD & 6 & 0.8702 & $1.68 \times 10^{-2}$ & 135 & 62 \\
Exact & 6 & 1.0000 & $0$ & 116 & 44 \\
\midrule
VDSP & 8 & 0.9997 & $7.56 \times 10^{-8}$ & 116 & 51 \\
PQC & 8 & 0.9996 & $1.33 \times 10^{-7}$ & 56 & 35 \\
MPD & 8 & 0.7554 & $5.94 \times 10^{-2}$ & 149 & 86 \\
Exact & 8 & 1.0000 & $0$ & 399 & 198 \\
\midrule
VDSP & 10 & 0.9898 & $1.04 \times 10^{-4}$ & 135 & 69 \\
PQC & 10 & 0.9780 & $4.86 \times 10^{-4}$ & 56 & 45 \\
MPD & 10 & 0.5631 & $1.84 \times 10^{-1}$ & 172 & 119 \\
Exact & 10 & 1.0000 & $0$ & 1680 & 873 \\
\midrule
VDSP & 12 & 0.9880 & $1.44 \times 10^{-4}$ & 150 & 82 \\
PQC & 12 & 0.9452 & $3.00 \times 10^{-3}$ & 56 & 55 \\
MPD & 12 & 0.5563 & $1.90 \times 10^{-1}$ & 190 & 141 \\
Exact & 12 & 1.0000 & $0$ & 6695 & 3612 \\
\midrule
VDSP & 14 & 0.9739 & $6.82 \times 10^{-4}$ & 159 & 95 \\
PQC & 14 & 0.2346 & $5.00 \times 10^{-1}$ & 56 & 65 \\
MPD & 14 & 0.6196 & $1.41 \times 10^{-1}$ & 205 & 168 \\
Exact & 14 & 1.0000 & $0$ & 27056 & 14762 \\
\midrule
VDSP & 16 & 0.9460 & $2.91 \times 10^{-3}$ & 179 & 113 \\
PQC & 16 & 0.2344 & $5.00 \times 10^{-1}$ & 56 & 75 \\
MPD & 16 & 0.6868 & $9.58 \times 10^{-2}$ & 214 & 191 \\
Exact & 16 & 1.0000 & $0$ & 108159 & 59517 \\
\midrule
VDSP & 18 & 0.9213 & $6.19 \times 10^{-3}$ & 195 & 125 \\
PQC & 18 & 0.2345 & $5.00 \times 10^{-1}$ & 56 & 85 \\
MPD & 18 & 0.6111 & $1.47 \times 10^{-1}$ & 239 & 215 \\
Exact & 18 & 1.0000 & $0$ & 433771 & 239305 \\
\midrule
VDSP & 20 & 0.9687 & $9.77 \times 10^{-4}$ & 211 & 141 \\
PQC & 20 & 0.2345 & $5.00 \times 10^{-1}$ & 56 & 95 \\
MPD & 20 & 0.4914 & $2.45 \times 10^{-1}$ & 252 & 250 \\
Exact & 20 & 1.0000 & $0$ & 1735318 & 959010 \\
\bottomrule
\end{tabular}
\caption{State preparation benchmark for the 2D Ricker wavelet distribution ($L_\mathrm{var}=5$, $L_\mathrm{MPS}=5$.)}
\label{tab:2d-ricker-main}
\end{table}

The results for the 2D normal distribution comparing VDSP and PQC with 2 variational layers $L_\mathrm{var}=2$ vs MPD with two disentangling layers $L_\mathrm{MPS} = 2$ and vs exact preparation is presented in Table \ref{tab:2d-normal-main}. Note that VDSP always has a single MPS layer. The key performance metrics are accuracy, infidelity, circuit depth and the number of CX of the transpiled state preparation circuit. Table \ref{tab:2d-ricker-main} collects the results for 2D Ricker wavelet for which we used 5 variational layers and a single MPD layer. We can see that for the 2D normal distribution the VDSP method consistently has an infidelity less than $10^{-4}$ in accuracy, beating the pure PQC and MPD methods in fidelity against the target state while using a comparable number of gates. The 2D Ricker wavelet has significantly higher entanglement entropy compared to the 2D normal distribution. For this state the VDSP reached the highest level of accuracy, with infidelity typically around $10^{-3}$. Meanwhile the PQC and MPD methods fail to accurately represent this state when the qubit number is increased beyond 12. In fact, for the number of qubits in the range 14--20, the VDSP is the only method besides exact state preparation that reaches a high fidelity against the target state.

Finally, we studied how the accuracy and infidelity improve when the number of layers is increased in Fig. \ref{fig:layer-scaling-2d-ricker}. We studied the 2D Ricker wavelet on 10 and 20 qubits and increased the number of layers from one to up to five. For a single layer on 10 qubits we observed that VDSP resulted in similar accuracy as a single PQC or MPD layer. For 2 and more layers the VDSP method had the best performance, although the accuracy started to stagnate after 4 layers. The scaling on 20 qubits was very different, when the PQC was unable to obtain an infidelity higher than 0.5, and the MPD showed very poor improvements in accuracy when increasing from 1 to 5 layers. Meanwhile, the VDSP was the only method which reached a decent infielity against the target state and showed increasing performance with increasing number of layers. This shows that minimizing entanglement entropy is a highly effective approach for the 2D Ricker wavelet.

\begin{figure*}
\centering
\pgfplotsset{every axis/.append style={
    line width=0.5pt,
}}
\begin{minipage}[t]{0.48\textwidth}
\centering
\begin{tikzpicture}
\begin{axis}[
    axis y line*=left,
    axis x line=bottom,
    xlabel={Number of layers},
    ylabel={Accuracy},
    title={$n_q = 10$},
    xmin=0.7, xmax=5.3,
    ymin=0.1, ymax=1.01,
    legend to name=sharedlegend,
    legend style={legend columns=3, font=\small, draw=black},
    width=0.95\linewidth
]
\addplot[color=tolred, mark=diamond*, mark options={solid, scale=1}] coordinates {
    (1, 0.527423)
    (2, 0.894537)
    (3, 0.974485)
    (4, 0.989504)
    (5, 0.989802)
};
\addlegendentry{VDSP (Accuracy)}
\addplot[color=tolgreen, mark=o, mark options={solid, scale=1}] coordinates {
    (1, 0.125819)
    (2, 0.479565)
    (3, 0.709115)
    (4, 0.959112)
    (5, 0.977950)
};
\addlegendentry{PQC (Accuracy)}
\addplot[color=tolblue, mark=x, mark options={solid, scale=1}] coordinates {
    (1, 0.226193)
    (2, 0.500110)
    (3, 0.553081)
    (4, 0.686662)
    (5, 0.563060)
};
\addlegendentry{MPD (Accuracy)}
\addlegendimage{color=tolred, dashed, mark=diamond*, mark options={solid, scale=1}}
\addlegendentry{VDSP (Infidelity)}
\addlegendimage{color=tolgreen, dashed, mark=o, mark options={solid, scale=1}}
\addlegendentry{PQC (Infidelity)}
\addlegendimage{color=tolblue, dashed, mark=x, mark options={solid, scale=1}}
\addlegendentry{MPD (Infidelity)}
\end{axis}

\begin{axis}[
    axis y line*=right,
    axis x line=none,
    ylabel={Infidelity},
    ymode=log,
    log basis y=10,
    ymin=1e-5, ymax=1e0,
    width=0.95\linewidth
]
\addplot[color=tolred, dashed, mark=diamond*, mark options={solid, scale=1}] coordinates {
    (1, 2.11e-01)
    (2, 1.11e-02)
    (3, 6.51e-04)
    (4, 1.10e-04)
    (5, 1.04e-04)
};
\addplot[color=tolgreen, dashed, mark=o, mark options={solid, scale=1}] coordinates {
    (1, 6.18e-01)
    (2, 2.53e-01)
    (3, 8.28e-02)
    (4, 1.67e-03)
    (5, 4.86e-04)
};
\addplot[color=tolblue, dashed, mark=x, mark options={solid, scale=1}] coordinates {
    (1, 5.09e-01)
    (2, 2.39e-01)
    (3, 1.90e-01)
    (4, 9.68e-02)
    (5, 1.84e-01)
};
\end{axis}
\end{tikzpicture}
\end{minipage}%
\hfill
\begin{minipage}[t]{0.48\textwidth}
\centering
\begin{tikzpicture}
\begin{axis}[
    axis y line*=left,
    axis x line=bottom,
    xlabel={Number of layers},
    ylabel={Accuracy},
    title={$n_q = 20$},
    xmin=0.7, xmax=5.3,
    ymin=0.0, ymax=1.01,
    width=0.95\linewidth
]
\addplot[color=tolred, mark=diamond*, mark options={solid, scale=1}] coordinates {
    (1, 0.600480)
    (2, 0.850227)
    (3, 0.913813)
    (4, 0.948505)
    (5, 0.968735)
};
\addplot[color=tolgreen, mark=o, mark options={solid, scale=1}] coordinates {
    (1, 0.000000)
    (2, 0.227393)
    (3, 0.233188)
    (4, 0.234282)
    (5, 0.234515)
};
\addplot[color=tolblue, mark=x, mark options={solid, scale=1}] coordinates {
    (1, 0.208590)
    (2, 0.444305)
    (3, 0.385722)
    (4, 0.555557)
    (5, 0.491415)
};
\end{axis}

\begin{axis}[
    axis y line*=right,
    axis x line=none,
    ylabel={Infidelity},
    ymode=log,
    log basis y=10,
    ymin=1e-5, ymax=1e0,
    width=0.95\linewidth
]
\addplot[color=tolred, dashed, mark=diamond*, mark options={solid, scale=1}] coordinates {
    (1, 1.53e-01)
    (2, 2.23e-02)
    (3, 7.41e-03)
    (4, 2.65e-03)
    (5, 9.77e-04)
};
\addplot[color=tolgreen, dashed, mark=o, mark options={solid, scale=1}] coordinates {
    (1, 7.52e-01)
    (2, 5.08e-01)
    (3, 5.02e-01)
    (4, 5.00e-01)
    (5, 5.00e-01)
};
\addplot[color=tolblue, dashed, mark=x, mark options={solid, scale=1}] coordinates {
    (1, 5.28e-01)
    (2, 2.90e-01)
    (3, 3.44e-01)
    (4, 1.91e-01)
    (5, 2.45e-01)
};
\end{axis}
\end{tikzpicture}
\end{minipage}

\vspace{0.5em}
\ref*{sharedlegend}
    \caption{Scaling of accuracy and infidelity with respect to number of layers for the 2D Ricker wavelet distribution. Left: $n_q = 10$. Right: $n_q = 20$. For VDSP and PQC the number of layers refers to the number of variational layers, while for MPD it refers to the number of MPD layers.}
    \label{fig:layer-scaling-2d-ricker}
\end{figure*}

In the benchmarks we used \texttt{qiskit} version 2.3.0 with the default gate set of \texttt{GenericBackendV2} \footnote{The default gate set consists of $\id$, $R_z$, $SX$, $X$ and $CX$ gates.}  for transpiling quantum circuits as well as for exact state preparation \cite{Plesch2011Quantum}, while the PQCs were trained with \texttt{quimb} \cite{Gray2018}. The PQCs were trained with the \texttt{L-BFGS-B} optimizer for 5000 rounds initially, followed by additional training with the \texttt{adam} optimizer until the training progress stalled. We allowed for maximum 30000 training rounds, however typically around 10000 rounds were sufficient. 

All of the PQCs were trained with a MacBook Pro with the M4 Pro processor and 48 GB of RAM. We found that the total memory consumption reached at most around 150 MB, showing that memory consumption was not the main bottleneck in our benchmarks for this number of qubits. The training time of the VDSP reached around 1 hour for 20 qubits. In the Section~\ref{sec:discussion} we discuss possible ways of scaling the VDSP method beyond the results presented here.

\subsection{Benchmark analysis}\label{section:benchmark-analysis}

Let us take a closer look at the benchmark states and the properties of the states produced by the different methods. Tables \ref{tab:bond_dim_2d_normal} and \ref{tab:bond_dim_2d_ricker} collect the maximum bond dimensions of the 2D normal distribution and 2D Ricker wavelet states, along with the maximum bond dimensions of the output states produced by the different methods. In these tables we used a threshold cutoff value of $1e^{-10}$ for the singular values.

In Section \ref{sec:theory} we introduced two lower bounds on the number of required variational layers for the VDSP method. Proposition \ref{prop:brickwall-layers} is a lower bound on exact state preparation, while Proposition \ref{prop:rank-reduction} is based on rank reduction. In the 20 qubit case these propositions give the numbers 30 and 2 for the 2D normal distribution, while for the 2D Ricker wavelet these numbers are 70, 3, respectively. Generally speaking we can say that the lower bound of Proposition \ref{prop:brickwall-layers} seems to be on the high side for these specific distributions, while the lower bound of Proposition \ref{prop:rank-reduction} seems to be more reasonable. However, we can immediately say that rank reduction is not the main mechanism by which the VDSP method works.

\begin{table}[t]
  \centering
  \begin{tabular}{r r r r r r}
    \toprule
    $n$ & Target & VDSP & PQC & MPD & Exact \\
    \midrule
    6 & 4 & 8 & 8 & 8 & 4 \\
    8 & 4 & 14 & 13 & 16 & 4 \\
    10 & 8 & 17 & 16 & 32 & 8 \\
    12 & 8 & 20 & 16 & 32 & 8 \\
    14 & 11 & 30 & 20 & 32 & 11 \\
    16 & 12 & 21 & 19 & 32 & 12 \\
    18 & 12 & 22 & 20 & 32 & 12 \\
    20 & 12 & 30 & 20 & 32 & 12 \\
    \bottomrule
  \end{tabular}
    \caption{Maximum bond dimension of target state and method output states for the 2D Ricker wavelet distribution ($L = 5$).}
  \label{tab:bond_dim_2d_ricker}
\end{table}

\begin{table}[t]
  \centering
  \begin{tabular}{r r r r r r}
    \toprule
    $n$ & Target & VDSP & PQC & MPD & Exact \\
    \midrule
    6 & 6 & 7 & 4 & 4 & 6 \\
    8 & 7 & 8 & 4 & 4 & 7 \\
    10 & 8 & 8 & 4 & 4 & 8 \\
    12 & 8 & 8 & 4 & 4 & 8 \\
    14 & 8 & 8 & 4 & 4 & 8 \\
    16 & 8 & 8 & 4 & 4 & 8 \\
    18 & 8 & 8 & 4 & 4 & 8 \\
    20 & 8 & 8 & 4 & 4 & 8 \\
    \bottomrule
  \end{tabular}
    \caption{Maximum bond dimension of target state and method output states for the 2D normal distribution ($L = 2$).}
  \label{tab:bond_dim_2d_normal}
\end{table}

In Table \ref{tab:bond_dim_2d_normal} it is clear that the PQC and MPD methods saturate the maximum bond dimension that is possible given the observations in Appendix \ref{appendix:mpo}; the methods are not able to produce an output state with higher bond dimension than 4. The VDSP method is able to produce a state with bond dimension 8 with two variational layers, which matches the bond dimension of the target state for 10--20 qubits.

The situation is drastically different for the 2D Ricker wavelet state, where we used 5 parametrized layers for VDSP and PQC and 5 disentangling layers for MPD. We can see that the MPD method produces an output state with the highest bond dimension, although the accuracy of this method was not great. The VDSP method had the best accuracy, but we see that the bond dimension of the output state is much higher than the bond dimension of the target state for all numbers of qubits. This must mean that the VDSP method does not fundamentally work by reducing the number of parameters or reducing the bond dimension of the target state. Instead, the VDSP method must work by redistributing the Schmidt mass of the target state with carefully picked rotations.

Theorem \ref{theorem:interference} explains the mechanism by which entanglement entropy is reduced by unitary gates. It could be possible to design an efficient greedy algorithm based on that mechanism. The VDSP uses global optimization instead. We can study how each CX gate reduces entanglement entropy in the trained ansatz. For this example we choose the 10-qubit 2D Ricker wavelet state with an ansatz consisting of 5 layers. Reading from Table \ref{tab:appendix-2d-ricker-10q}, the trained ansatz is able to reduce entanglement entropy from $2.179$ to $0.0038$, a roughly 98 \% reduction. Figure \ref{fig:entanglement-reduction-graph} shows how the cumulative entanglement entropy \eqref{eq:ee-loss} changes when applying the trained ansatz to the target state one CX gate at a time. We see that entanglement entropy does not monotonically decrease as there are multiple CX gates that increase the entanglement entropy.

Figure \ref{fig:ee-heatmap} provides another visualization for how the entanglement entropy is reduced for the 10-qubit 2D Ricker wavelet state. Here the CX gates are visualized in a 2D grid, and the entanglement entropy reduction after each CX gate is presented as a heatmap. We observe significant redistribution of entanglement before the final layer after which entanglement entropy becomes very low. The maximum bond dimension for this state is $8$ by Table \ref{tab:bond_dim_2d_ricker}, so especially the bonds with high entanglement entropies display entanglement across more than two qubits. Nevertheless the VDSP method is very effective at reducing the entanglement entropy.

\begin{figure}
    \centering

\begin{tikzpicture}
  \begin{axis}[
    width=\columnwidth,
    height=0.62\columnwidth,
    xmin=-0.5, xmax=44.5,
    ymin=-0.15,
    xlabel={CX gate index},
    ylabel={$S$},
    label style={font=\small},
    tick label style={font=\scriptsize},
    xtick={0,5,10,15,20,25,30,35,40,44},
    legend style={at={(0.98,0.98)}, anchor=north east, font=\scriptsize},
    clip=false,
  ]
    \addplot[gray, dashed, thick, forget plot]
      coordinates {(-0.5,2.178854e+00) (44.5,2.178854e+00)};
    \addlegendimage{gray, dashed, thick}
    \addlegendentry{Before ansatz}

    \addplot[gray, dotted, thick, forget plot]
      coordinates {(-0.5,3.758082e-03) (44.5,3.758082e-03)};
    \addlegendimage{gray, dotted, thick}
    \addlegendentry{After ansatz}

    \addplot[blue!70!black, const plot mark mid, thick] coordinates {
      (0,2.17885450e+00) (1,2.18682115e+00) (2,3.09353190e+00)
      (3,3.06380048e+00) (4,3.06380048e+00) (5,3.06385811e+00)
      (6,3.52818788e+00) (7,2.60856170e+00) (8,2.61756102e+00)
      (9,2.61756102e+00) (10,2.60482229e+00) (11,1.69811168e+00)
      (12,1.34575573e+00) (13,1.34575573e+00) (14,1.34391988e+00)
      (15,6.40179850e-01) (16,6.38870758e-01) (17,8.43822950e-01)
      (18,8.43822950e-01) (19,7.40454305e-01) (20,7.77823344e-01)
      (21,7.06561923e-01) (22,7.06561923e-01) (23,7.07377647e-01)
      (24,6.88609747e-01) (25,7.15218103e-01) (26,6.30975129e-01)
      (27,6.30975129e-01) (28,6.66776947e-01) (29,6.49184362e-01)
      (30,6.52391130e-01) (31,6.52391130e-01) (32,6.52187486e-01)
      (33,6.57146445e-01) (34,6.31105604e-01) (35,8.50560963e-01)
      (36,8.50560963e-01) (37,4.77466626e-01) (38,4.57771461e-01)
      (39,4.13789462e-01) (40,4.13789462e-01) (41,4.11214569e-01)
      (42,3.88973135e-01) (43,3.85267033e-01) (44,3.75808171e-03)
    };
    \addplot[blue!70!black, only marks, mark=*, mark size=1.5pt] coordinates {
      (0,2.17885450e+00) (1,2.18682115e+00) (2,3.09353190e+00)
      (3,3.06380048e+00) (4,3.06380048e+00) (5,3.06385811e+00)
      (6,3.52818788e+00) (7,2.60856170e+00) (8,2.61756102e+00)
      (9,2.61756102e+00) (10,2.60482229e+00) (11,1.69811168e+00)
      (12,1.34575573e+00) (13,1.34575573e+00) (14,1.34391988e+00)
      (15,6.40179850e-01) (16,6.38870758e-01) (17,8.43822950e-01)
      (18,8.43822950e-01) (19,7.40454305e-01) (20,7.77823344e-01)
      (21,7.06561923e-01) (22,7.06561923e-01) (23,7.07377647e-01)
      (24,6.88609747e-01) (25,7.15218103e-01) (26,6.30975129e-01)
      (27,6.30975129e-01) (28,6.66776947e-01) (29,6.49184362e-01)
      (30,6.52391130e-01) (31,6.52391130e-01) (32,6.52187486e-01)
      (33,6.57146445e-01) (34,6.31105604e-01) (35,8.50560963e-01)
      (36,8.50560963e-01) (37,4.77466626e-01) (38,4.57771461e-01)
      (39,4.13789462e-01) (40,4.13789462e-01) (41,4.11214569e-01)
      (42,3.88973135e-01) (43,3.85267033e-01) (44,3.75808171e-03)
    };
  \end{axis}
\end{tikzpicture}
    \caption{Gate by gate reduction in cumulative entanglement entropy by a trained 5-layer PQC for the 10-qubit 2D Ricker wavelet state.}
    \label{fig:entanglement-reduction-graph}
\end{figure}

\begin{figure}
    \centering
    \input{figures/cx_ee_panel_b}
    \caption{Heatmap of entanglement entropy reduction for the 10-qubit 2D Ricker wavelet state. Each red cross represents a CX gate in the trained ansatz acting on bond $i$. A brick wall PQC is organized into even and odd layers and this clarified on the horizontal axis.}
    \label{fig:ee-heatmap}
\end{figure}

\subsection{Noise analysis}

In the NISQ era it is important to analyze the effect of noise. Roughly speaking, the fidelity loss can be estimated by taking a product of individual gate fidelities. Each gate fidelity should be raised to the power of the respective gates occurrence in the circuit. However, this simple model fails to take coherence times, readout errors and crosstalk, among other noise sources, into consideration. 

To analyze the noise in a consistent way, we create a custom noise model in \texttt{qiskit} with varying levels of gate fidelities. We assume a depolarizing noise channel for one- and two-qubit gates as an appropriate model for average noise in large circuits \cite{Urbanek2021Mitigating}, while readout error is modelled by symmetric bit-flip measurement error. Specifically, we analyze two-qubit gate fidelities at six different levels: 0.99, 0.995, 0.999, 0.9995, 0.9999, 0.99999, one-qubit gate fidelity is set to 0.99999, while readout errors are analyzed at three different levels: 0.001, 0.005, 0.01. These noise ranges represent a realistic model for current or near-term generation superconducting \cite{abdurakhimov2024technologyperformancebenchmarksiqms} and ion devices \cite{ransford2025helios98qubittrappedionquantum}.

To compare the outputs of circuits we adopt a simple distance metric with the Hellinger distance:\begin{equation}
    H\left( P , Q \right) = \sqrt{1 - \sum_i \sqrt{p_i q_i} }.
\end{equation} The Hellinger distance compares the squared amplitudes of the target state to the observed probability distribution over the outcomes of a computational basis measurement. This forms a metric in the set of probability distributions and avoids comparing density matrices directly and any complications with full state tomography. A better circuit will have a lower Hellinger distance to the target state. 

In the name of transparency, it makes a big difference how we choose the number of layers in the noise comparison. There is a tradeoff between the number of layers and ideal (zero noise) fidelity ($F_\mathrm{ideal}$ in Table \ref{tab:noise-benchmark}) against the target state. The ideal fidelity of the circuit then has to be compared to the gate fidelities for the given noise model the circuit is run on. We will show an example noise analysis with the 2D Ricker wavelet state prepared on 10 qubits. We compare the VDSP with three variational layers to the PQC and MPD circuits with three layers. The comparison is presented in Fig. \ref{fig:hellinger-panel}. We note that this comparison is not entirely fair to the PQC method, as the PQC with 5 layers reaches a comparable fidelity compared to VDSP with 3 layers with similar gate counts. The full noise simulation is presented in Table \ref{tab:noise-benchmark}. However, additional layers fail to make the PQC more accurate starting from 14 qubits. Unfortunately, running the noise models with 14 qubits is computationally expensive and it was not possible on the computer we were using. It is clear that with 14-20 qubits the VDSP method provides the only viable method in our comparison for preparing the 2D Ricker wavelet state on NISQ devices. Even with 10 qubits, when the PQC offers good ideal fidelity to the target state, the VDSP method obtains the best Hellinger distance of $0.06$ at CZ gate fidelity $0.9999$, while the PQC only obtains this Hellinger distance at the final CZ fidelity $0.99999$. Meanwhile, the MPD method was unable to provide a good quality circuit, and performance of the method starts to deteriorate after 4 layers.

\begin{table*}[t]
  \centering
  \begin{tabular}{llrrrrrrrrrr}
    \toprule
    Method & $L$ & CX & Depth & $F_{\mathrm{ideal}}$ & $H_{\mathrm{ideal}}$ & $H_{0.99}$ & $H_{0.995}$ & $H_{0.999}$ & $H_{0.9995}$ & $H_{0.9999}$ & $H_{0.99999}$ \\
    \midrule
    VDSP & 1 & 33 & 93 & $7.89 \times 10^{-1}$ & 0.30 & 0.34 & 0.31 & 0.30 & 0.30 & 0.30 & 0.30 \\
    VDSP & 2 & 38 & 101 & $9.89 \times 10^{-1}$ & 0.07 & 0.23 & 0.17 & 0.11 & 0.10 & 0.09 & 0.09 \\
    VDSP & 3 & 49 & 110 & $9.99 \times 10^{-1}$ & 0.02 & 0.25 & 0.18 & 0.09 & 0.08 & 0.06 & 0.06 \\
    VDSP & 4 & 60 & 123 & 1.00 & 0.01 & 0.28 & 0.20 & 0.10 & 0.08 & 0.06 & 0.06 \\
    VDSP & 5 & 69 & 135 & 1.00 & 0.01 & 0.31 & 0.22 & 0.11 & 0.09 & 0.07 & 0.07 \\
    \midrule
    PQC & 1 & 9 & 16 & $3.82 \times 10^{-1}$ & 0.54 & 0.51 & 0.52 & 0.53 & 0.53 & 0.53 & 0.53 \\
    PQC & 2 & 18 & 26 & $7.47 \times 10^{-1}$ & 0.33 & 0.35 & 0.34 & 0.33 & 0.33 & 0.33 & 0.33 \\
    PQC & 3 & 27 & 36 & $9.17 \times 10^{-1}$ & 0.20 & 0.27 & 0.23 & 0.21 & 0.21 & 0.21 & 0.21 \\
    PQC & 4 & 36 & 46 & $9.98 \times 10^{-1}$ & 0.03 & 0.23 & 0.16 & 0.09 & 0.08 & 0.07 & 0.06 \\
    PQC & 5 & 45 & 56 & 1.00 & 0.02 & 0.27 & 0.19 & 0.10 & 0.08 & 0.07 & 0.06 \\
    \midrule
    MPD & 1 & 25 & 85 & $4.91 \times 10^{-1}$ & 0.45 & 0.44 & 0.44 & 0.45 & 0.45 & 0.45 & 0.45 \\
    MPD & 2 & 47 & 106 & $7.61 \times 10^{-1}$ & 0.33 & 0.37 & 0.34 & 0.33 & 0.33 & 0.33 & 0.33 \\
    MPD & 3 & 69 & 129 & $8.10 \times 10^{-1}$ & 0.31 & 0.39 & 0.35 & 0.31 & 0.31 & 0.31 & 0.31 \\
    MPD & 4 & 96 & 149 & $9.03 \times 10^{-1}$ & 0.22 & 0.37 & 0.30 & 0.23 & 0.23 & 0.22 & 0.22 \\
    MPD & 5 & 119 & 172 & $8.16 \times 10^{-1}$ & 0.30 & 0.49 & 0.40 & 0.32 & 0.31 & 0.30 & 0.30 \\
    \midrule
    Exact & -- & 873 & 1680 & 1 & 0.00 & 0.78 & 0.74 & 0.48 & 0.37 & 0.25 & 0.20 \\
    \bottomrule
  \end{tabular}
  \caption{%
    Noise simulation results for the 10--qubit 2D Ricker wavelet state.
    Hellinger distance ($H$) is reported at representative CX gate
    fidelities with single-qubit fidelity $F_{1Q} = 0.99999$ and readout
    error rate $\varepsilon_{\mathrm{ro}} = 0.001$.
  }\label{tab:noise-benchmark}
\end{table*}


\begin{figure*}[t]
  \centering
  \begin{tikzpicture}
    \begin{groupplot}[
      group style={
        group size=3 by 1,
        horizontal sep=0.15cm,
      },
      width=0.36\textwidth,
      height=0.28\textwidth,
      xmode=log,
      ymode=log,
      x dir=reverse,
      ymin=0.04,
      ymax=1.0,
      ytick={0.1, 0.2, 0.3, 0.5, 0.8},
      yticklabels={0.1, 0.2, 0.3, 0.5, 0.8},
      xlabel={CZ gate infidelity ($1{-}F$)},
      grid=both,
      grid style={line width=0.2pt, draw=gray!30},
      major grid style={line width=0.4pt, draw=gray!50},
      legend style={
        font=\scriptsize,
        at={(0.98,0.98)},
        anchor=north east,
        draw=none,
        fill=white,
        fill opacity=0.8,
        text opacity=1,
      },
      every axis title/.style={font=\footnotesize, at={(0.5,1)}, anchor=south},
      tick label style={font=\scriptsize},
      label style={font=\footnotesize},
      cycle list name=exotic,
    ]

    \nextgroupplot[
      title={$\varepsilon_{\mathrm{ro}} = 0.001$},
      ylabel={Hellinger distance},
    ]
    \addplot coordinates {(1.0e-02,0.250014) (5.0e-03,0.176928) (1.0e-03,0.090140) (5.0e-04,0.075579) (1.0e-04,0.063222) (1.0e-05,0.060367)};
    \addlegendentry{VDSP}
    \addplot coordinates {(1.0e-02,0.265960) (5.0e-03,0.232340) (1.0e-03,0.211001) (5.0e-04,0.208595) (1.0e-04,0.207376) (1.0e-05,0.207502)};
    \addlegendentry{PQC}
    \addplot coordinates {(1.0e-02,0.390574) (5.0e-03,0.346534) (1.0e-03,0.312464) (5.0e-04,0.309672) (1.0e-04,0.309344) (1.0e-05,0.308711)};
    \addlegendentry{MDP}
    \addplot coordinates {(1.0e-02,0.782972) (5.0e-03,0.739878) (1.0e-03,0.477070) (5.0e-04,0.373351) (1.0e-04,0.246523) (1.0e-05,0.204173)};
    \addlegendentry{Exact}

    \nextgroupplot[
      title={$\varepsilon_{\mathrm{ro}} = 0.005$},
      yticklabels={},
    ]
    \addplot coordinates {(1.0e-02,0.265992) (5.0e-03,0.199117) (1.0e-03,0.126664) (5.0e-04,0.116104) (1.0e-04,0.108144) (1.0e-05,0.106092)};
    \addplot coordinates {(1.0e-02,0.279325) (5.0e-03,0.247151) (1.0e-03,0.225625) (5.0e-04,0.222710) (1.0e-04,0.221468) (1.0e-05,0.221370)};
    \addplot coordinates {(1.0e-02,0.399005) (5.0e-03,0.355715) (1.0e-03,0.321967) (5.0e-04,0.318648) (1.0e-04,0.317258) (1.0e-05,0.316637)};
    \addplot coordinates {(1.0e-02,0.783012) (5.0e-03,0.740268) (1.0e-03,0.483321) (5.0e-04,0.382729) (1.0e-04,0.262036) (1.0e-05,0.222804)};

    \nextgroupplot[
      title={$\varepsilon_{\mathrm{ro}} = 0.01$},
      yticklabels={},
    ]
    \addplot coordinates {(1.0e-02,0.285114) (5.0e-03,0.224265) (1.0e-03,0.162481) (5.0e-04,0.153903) (1.0e-04,0.147841) (1.0e-05,0.145998)};
    \addplot coordinates {(1.0e-02,0.295818) (5.0e-03,0.265168) (1.0e-03,0.243632) (5.0e-04,0.240653) (1.0e-04,0.239149) (1.0e-05,0.239164)};
    \addplot coordinates {(1.0e-02,0.410026) (5.0e-03,0.367685) (1.0e-03,0.334372) (5.0e-04,0.330371) (1.0e-04,0.328365) (1.0e-05,0.327503)};
    \addplot coordinates {(1.0e-02,0.782756) (5.0e-03,0.741287) (1.0e-03,0.490820) (5.0e-04,0.394474) (1.0e-04,0.280975) (1.0e-05,0.244838)};

    \end{groupplot}
  \end{tikzpicture}
  \caption{%
    Hellinger distance between the noisy simulator output and the target
    distribution as a function of CZ gate infidelity ($1 - F_{\mathrm{CZ}}$)
    for four state-preparation methods on an 10-qubit 2D Ricker wavelet state.
    Single-qubit gate fidelity is $F_{1Q} = 0.99999$.
    Panels correspond to readout error rates
    $\varepsilon_{\mathrm{ro}} \in \{0.001, 0.005, 0.01\}$.
  }
  \label{fig:hellinger-panel}
\end{figure*}

\subsection{Hardware results}

We chose the 2D Ricker wavelet on 6 qubits as the benchmark state for hardware tests. The entanglement entropy of this state is $0.4222$, much higher than the 2D normal distribution, so it represents an interesting case study. We used the VDSP method with one variational layer, PQC with three layers, MPD with one disentangling layer and compared these circuits to the exact method. The accuracy and infidelity against the target state for all of the methods can be found from Table \ref{tab:appendix-2d-ricker-6q} in the Appendix. The infidelities of the methods against the target state were $7.02 \times 10^{-4}$ for VDSP, $8.61 \times 10^{-7}$ for PQC and $1.54 \times 10^{-1}$ for MPD.

We tested all four state preparation methods on the VTT-IQM Q50 device, which is available at \url{https://qx.vtt.fi/}. The device has a flux tunable qubit and coupler architecture with a CZ and PRX native gateset \cite{iqm1,iqm2}, on a square lattice topology and is calibrated twice a day \footnote{At the time we ran the tests on 27th April 2026 the calibration set id was \texttt{21b3675b-52f1-48fe-9ef1-af7ee90d869a}.}. We provide the average gate fidelities and coherence times of the device in Table \ref{tab:calibration_summary}. 

We identified the chain of qubits Q23-Q24-Q16-Q17-Q25-Q26 as a subset of physical qubits with the best average CZ fidelity. The average CZ RB fidelity of these qubits was $0.994440$, while the average T1 and T2 (Echo) times were 34.7 $\mu s$ and 13.2 $\mu s$, respectively. However, the minimum T2 (Echo) time was 5.5 $\mu s$ for qubit Q23. This leads us to suspect that the circuit depth and CZ count contribute more to the quality of the state preparation than ideal fidelity against the target state.

Via the IQM client API, we applied digital dynamical decoupling (XX pulse scheme) as well as heralding post-selection on the readout states. In addition, we applied twirled readout error mitigation \cite{Berg2022Model, Smith2021Qubit, Hashim2025Quasiprobabilistic} as well as state-aware circuit randomization, where random bitflips are added to the existing single qubit gates according to a non-uniform probability distribution informed by the qubits magnetization \cite{iqm2026unpublished}. For each state-preparation circuit, we distributed the total number of shots over 20 random instances. The readout error mitigation was implemented via the M3 Qiskit addon. These methods had the largest improvement on the methods with higher circuit depth and thereby reduced the effect of decoherence on idle qubits.

The results are represented in Figure \ref{fig:iqm-combined}. The MPD method performs well, obtaining a Hellinger distance of 0.388, considering it had the lowest ideal fidelity against the target state. The PQC and VDSP methods are very close in performance, but the VDSP method benefits most from the readout twirling, and manages to edge out a close win with respect to the Hellinger distance. The VDSP obtained a Hellinger distance of 0.354 while for PQC the Hellinger distance was 0.363. These results highlight that, while it is possible to estimate the performance of different methods from theory and noise analysis, it is not enough to simply consider the CX gate count and simple noise sources. Error mitigation plays a significant role in the NISQ era, and ultimately the best method for state preparation has to be found by trial-and-error.


\begin{table}[htbp]
\centering

\small
\begin{tabular}{@{}lccc@{}}
\toprule
Metric & Median & Average \\ \midrule
$T_1$ Time [$\mu s$]      & 28.257 & 28.329 \\
$T_2$ Time (Ramsey) [$\mu s$]  & 6.974  & 8.423  \\
$T_2$ Time (Echo) [$\mu s$]  & 16.582 & 17.030  \\ \addlinespace
PRX RB Gate Fidelity [\%]  & 99.939 & 99.822  \\
Readout Fidelity [\%]   & 97.450 & 97.029  \\
CZ RB Gate Fidelity [\%]   & 98.878 & 97.952 \\
\bottomrule
\end{tabular}
\caption{Summary of device performance for calibration set \texttt{21b3675b-52f1-48fe-9ef1-af7ee90d869a}.}
\label{tab:calibration_summary}
\end{table}


\begin{figure*}[t]
  \centering
  \begin{tikzpicture}
    \pgfplotsset{
      target bar/.style={
        ybar, bar width=1.8pt, draw=none,
        fill=black!25, opacity=0.8,
      },
      vdsp bar/.style={
        ybar, bar width=1.8pt, draw=none,
        fill=blue!70, opacity=0.85,
      },
      pqc bar/.style={
        ybar, bar width=1.8pt, draw=none,
        fill=orange!80, opacity=0.85,
      },
      mpd bar/.style={
        ybar, bar width=1.8pt, draw=none,
        fill=green!60!black, opacity=0.85,
      },
      exact bar/.style={
        ybar, bar width=1.8pt, draw=none,
        fill=violet!70, opacity=0.85,
      },
    }
    \begin{groupplot}[
      group style={
        group size=2 by 2,
        horizontal sep=1.2cm,
        vertical sep=1.0cm,
      },
      area legend,
      width=0.50\textwidth,
      height=0.30\textwidth,
      xmin=-1, xmax=64,
      ymin=0,
      xtick={0,10,20,30,40,50,60},
      grid=both,
      grid style={line width=0.2pt, draw=gray!20},
      major grid style={line width=0.4pt, draw=gray!40},
      legend style={
        font=\scriptsize,
        at={(0.97,0.97)},
        anchor=north east,
        draw=none,
        fill=white,
        fill opacity=0.85,
        text opacity=1,
      },
      every axis title/.style={font=\small, at={(0.5,1)}, anchor=south},
      tick label style={font=\scriptsize},
      label style={font=\small},
    ]

    \nextgroupplot[
      title={VDSP ($H=0.354$, $N_{\mathrm{CZ}}=19$)},
      ylabel={Probability},
    ]
    \addplot[target bar] coordinates {(0,0.00000000) (1,0.00000000) (2,0.00000000) (3,0.00000000) (4,0.00000000) (5,0.00000000) (6,0.00000000) (7,0.00000000) (8,0.00000000) (9,0.00000000) (10,0.00000000) (11,0.00000000) (12,0.00000000) (13,0.00000000) (14,0.00000000) (15,0.00000000) (16,0.00000000) (17,0.00000000) (18,0.00000005) (19,0.00007301) (20,0.00072266) (21,0.00007301) (22,0.00000005) (23,0.00000000) (24,0.00000000) (25,0.00000000) (26,0.00007301) (27,0.03794625) (28,0.09294258) (29,0.03794625) (30,0.00007301) (31,0.00000000) (32,0.00000000) (33,0.00000000) (34,0.00072266) (35,0.09294258) (36,0.47296971) (37,0.09294258) (38,0.00072266) (39,0.00000000) (40,0.00000000) (41,0.00000000) (42,0.00007301) (43,0.03794625) (44,0.09294258) (45,0.03794625) (46,0.00007301) (47,0.00000000) (48,0.00000000) (49,0.00000000) (50,0.00000005) (51,0.00007301) (52,0.00072266) (53,0.00007301) (54,0.00000005) (55,0.00000000) (56,0.00000000) (57,0.00000000) (58,0.00000000) (59,0.00000000) (60,0.00000000) (61,0.00000000) (62,0.00000000) (63,0.00000000)};
    \addlegendentry{Target}
    \addplot[vdsp bar] coordinates {(0,0.00001551) (1,0.00000000) (2,0.00033441) (3,0.00000000) (4,0.00000000) (5,0.00000000) (6,0.00000000) (7,0.00000000) (8,0.00025437) (9,0.00010668) (10,0.00035740) (11,0.00076866) (12,0.00580234) (13,0.00237880) (14,0.00066653) (15,0.00031010) (16,0.00087988) (17,0.00040615) (18,0.00203988) (19,0.00622302) (20,0.01699268) (21,0.00516857) (22,0.00160145) (23,0.00423916) (24,0.00420404) (25,0.00441928) (26,0.00771250) (27,0.02482573) (28,0.06433213) (29,0.03076923) (30,0.00825855) (31,0.01088538) (32,0.00963951) (33,0.01132106) (34,0.02072853) (35,0.07733546) (36,0.36036798) (37,0.07708187) (38,0.03754870) (39,0.03548833) (40,0.00258814) (41,0.00438230) (42,0.00837211) (43,0.02104094) (44,0.05130977) (45,0.03048798) (46,0.00811544) (47,0.00921318) (48,0.00014040) (49,0.00029001) (50,0.00010526) (51,0.00170335) (52,0.00192348) (53,0.00228420) (54,0.00021435) (55,0.00100403) (56,0.00014876) (57,0.00034771) (58,0.00157860) (59,0.00355712) (60,0.01159891) (61,0.00318917) (62,0.00161566) (63,0.00132521)};
    \addlegendentry{VDSP}

    \nextgroupplot[
      title={PQC ($H=0.363$, $N_{\mathrm{CZ}}=15$)},
      yticklabels={},
    ]
    \addplot[target bar] coordinates {(0,0.00000000) (1,0.00000000) (2,0.00000000) (3,0.00000000) (4,0.00000000) (5,0.00000000) (6,0.00000000) (7,0.00000000) (8,0.00000000) (9,0.00000000) (10,0.00000000) (11,0.00000000) (12,0.00000000) (13,0.00000000) (14,0.00000000) (15,0.00000000) (16,0.00000000) (17,0.00000000) (18,0.00000005) (19,0.00007301) (20,0.00072266) (21,0.00007301) (22,0.00000005) (23,0.00000000) (24,0.00000000) (25,0.00000000) (26,0.00007301) (27,0.03794625) (28,0.09294258) (29,0.03794625) (30,0.00007301) (31,0.00000000) (32,0.00000000) (33,0.00000000) (34,0.00072266) (35,0.09294258) (36,0.47296971) (37,0.09294258) (38,0.00072266) (39,0.00000000) (40,0.00000000) (41,0.00000000) (42,0.00007301) (43,0.03794625) (44,0.09294258) (45,0.03794625) (46,0.00007301) (47,0.00000000) (48,0.00000000) (49,0.00000000) (50,0.00000005) (51,0.00007301) (52,0.00072266) (53,0.00007301) (54,0.00000005) (55,0.00000000) (56,0.00000000) (57,0.00000000) (58,0.00000000) (59,0.00000000) (60,0.00000000) (61,0.00000000) (62,0.00000000) (63,0.00000000)};
    \addlegendentry{Target}
    \addplot[pqc bar] coordinates {(0,0.00079642) (1,0.00080354) (2,0.00288823) (3,0.00376511) (4,0.02124148) (5,0.00286615) (6,0.00074097) (7,0.00066365) (8,0.00051669) (9,0.00020205) (10,0.00044965) (11,0.00177402) (12,0.00213494) (13,0.00090024) (14,0.00088153) (15,0.00008601) (16,0.00108628) (17,0.00093931) (18,0.00205654) (19,0.00462973) (20,0.02066761) (21,0.00390037) (22,0.00371659) (23,0.00133172) (24,0.00587596) (25,0.00164902) (26,0.00664679) (27,0.02142692) (28,0.06457260) (29,0.02222931) (30,0.00684336) (31,0.00150984) (32,0.01577802) (33,0.00751291) (34,0.02951846) (35,0.06938175) (36,0.40634590) (37,0.05695764) (38,0.00909103) (39,0.00850776) (40,0.00384074) (41,0.00207043) (42,0.00462088) (43,0.01892155) (44,0.05107863) (45,0.02090736) (46,0.00707714) (47,0.00159010) (48,0.00150210) (49,0.00107868) (50,0.00184985) (51,0.00232415) (52,0.00805470) (53,0.00137757) (54,0.00401809) (55,0.00269973) (56,0.00252789) (57,0.00076471) (58,0.00282756) (59,0.00798253) (60,0.02568379) (61,0.00944608) (62,0.00362071) (63,0.00124897)};
    \addlegendentry{PQC}

    \nextgroupplot[
      title={MPD ($H=0.388$, $N_{\mathrm{CZ}}=13$)},
      ylabel={Probability},
      xlabel={Basis state},
    ]
    \addplot[target bar] coordinates {(0,0.00000000) (1,0.00000000) (2,0.00000000) (3,0.00000000) (4,0.00000000) (5,0.00000000) (6,0.00000000) (7,0.00000000) (8,0.00000000) (9,0.00000000) (10,0.00000000) (11,0.00000000) (12,0.00000000) (13,0.00000000) (14,0.00000000) (15,0.00000000) (16,0.00000000) (17,0.00000000) (18,0.00000005) (19,0.00007301) (20,0.00072266) (21,0.00007301) (22,0.00000005) (23,0.00000000) (24,0.00000000) (25,0.00000000) (26,0.00007301) (27,0.03794625) (28,0.09294258) (29,0.03794625) (30,0.00007301) (31,0.00000000) (32,0.00000000) (33,0.00000000) (34,0.00072266) (35,0.09294258) (36,0.47296971) (37,0.09294258) (38,0.00072266) (39,0.00000000) (40,0.00000000) (41,0.00000000) (42,0.00007301) (43,0.03794625) (44,0.09294258) (45,0.03794625) (46,0.00007301) (47,0.00000000) (48,0.00000000) (49,0.00000000) (50,0.00000005) (51,0.00007301) (52,0.00072266) (53,0.00007301) (54,0.00000005) (55,0.00000000) (56,0.00000000) (57,0.00000000) (58,0.00000000) (59,0.00000000) (60,0.00000000) (61,0.00000000) (62,0.00000000) (63,0.00000000)};
    \addlegendentry{Target}
    \addplot[mpd bar] coordinates {(0,0.00107022) (1,0.00000000) (2,0.00046570) (3,0.00283820) (4,0.01522197) (5,0.00050802) (6,0.00000000) (7,0.00000000) (8,0.00007654) (9,0.00019782) (10,0.00000000) (11,0.00194153) (12,0.00329571) (13,0.00000000) (14,0.00005845) (15,0.00042865) (16,0.00334333) (17,0.00170212) (18,0.00331105) (19,0.01475254) (20,0.03689647) (21,0.00174293) (22,0.00026921) (23,0.00172760) (24,0.00965179) (25,0.00367894) (26,0.01014586) (27,0.04335624) (28,0.06525607) (29,0.00277962) (30,0.00000000) (31,0.00488997) (32,0.01342942) (33,0.00616184) (34,0.01519699) (35,0.07906845) (36,0.45500916) (37,0.01566036) (38,0.00000000) (39,0.01051948) (40,0.00866586) (41,0.00312513) (42,0.00802251) (43,0.03447982) (44,0.05362872) (45,0.00235689) (46,0.00000000) (47,0.00326106) (48,0.00341459) (49,0.00075341) (50,0.00309982) (51,0.01435450) (52,0.02202039) (53,0.00076028) (54,0.00012339) (55,0.00127397) (56,0.00140487) (57,0.00064658) (58,0.00192713) (59,0.00552429) (60,0.00542518) (61,0.00031013) (62,0.00000000) (63,0.00076923)};
    \addlegendentry{MPD}

    \nextgroupplot[
      title={Exact ($H=0.761$, $N_{\mathrm{CZ}}=69$)},
      xlabel={Basis state},
      yticklabels={},
    ]
    \addplot[target bar] coordinates {(0,0.00000000) (1,0.00000000) (2,0.00000000) (3,0.00000000) (4,0.00000000) (5,0.00000000) (6,0.00000000) (7,0.00000000) (8,0.00000000) (9,0.00000000) (10,0.00000000) (11,0.00000000) (12,0.00000000) (13,0.00000000) (14,0.00000000) (15,0.00000000) (16,0.00000000) (17,0.00000000) (18,0.00000005) (19,0.00007301) (20,0.00072266) (21,0.00007301) (22,0.00000005) (23,0.00000000) (24,0.00000000) (25,0.00000000) (26,0.00007301) (27,0.03794625) (28,0.09294258) (29,0.03794625) (30,0.00007301) (31,0.00000000) (32,0.00000000) (33,0.00000000) (34,0.00072266) (35,0.09294258) (36,0.47296971) (37,0.09294258) (38,0.00072266) (39,0.00000000) (40,0.00000000) (41,0.00000000) (42,0.00007301) (43,0.03794625) (44,0.09294258) (45,0.03794625) (46,0.00007301) (47,0.00000000) (48,0.00000000) (49,0.00000000) (50,0.00000005) (51,0.00007301) (52,0.00072266) (53,0.00007301) (54,0.00000005) (55,0.00000000) (56,0.00000000) (57,0.00000000) (58,0.00000000) (59,0.00000000) (60,0.00000000) (61,0.00000000) (62,0.00000000) (63,0.00000000)};
    \addlegendentry{Target}
    \addplot[exact bar] coordinates {(0,0.01068415) (1,0.01335954) (2,0.01183919) (3,0.01561348) (4,0.01793961) (5,0.01150836) (6,0.01403220) (7,0.01390321) (8,0.01190394) (9,0.01433498) (10,0.01342457) (11,0.01447791) (12,0.01713241) (13,0.01203366) (14,0.01316528) (15,0.01470149) (16,0.01191398) (17,0.01353500) (18,0.01473864) (19,0.01742920) (20,0.01690933) (21,0.01136268) (22,0.01550191) (23,0.01200696) (24,0.01219979) (25,0.01254170) (26,0.01601203) (27,0.01847227) (28,0.01878750) (29,0.01168085) (30,0.01564549) (31,0.01540188) (32,0.01976731) (33,0.02266722) (34,0.02033832) (35,0.02429852) (36,0.03534653) (37,0.01765360) (38,0.01995973) (39,0.02137325) (40,0.01201591) (41,0.01703571) (42,0.01532070) (43,0.01827415) (44,0.02002371) (45,0.01228832) (46,0.01699506) (47,0.01600241) (48,0.01147645) (49,0.01409651) (50,0.01251217) (51,0.01917071) (52,0.01938574) (53,0.01379623) (54,0.01369290) (55,0.01429376) (56,0.01168567) (57,0.01497776) (58,0.01468869) (59,0.01529062) (60,0.01848735) (61,0.01327645) (62,0.01384393) (63,0.01577148)};
    \addlegendentry{Exact}

    \end{groupplot}
  \end{tikzpicture}
  \caption{%
  Probability distributions from 6-qubit 2D Ricker wavelet state
  preparation on the VTT-IQM Q50 device (20000 shots, heralding + DD + M3 mitigation).
  $H$ denotes the Hellinger distance to the target distribution.
}
  \label{fig:iqm-combined}
\end{figure*}

\section{Discussion}\label{sec:discussion}

We have introduced an innovative, NISQ friendly method for quantum state preparation. We demonstrated the efficacy and accuracy of the VDSP method by using it to prepare 2D normal distribution and Ricker wavelet states. Our method demonstrated the highest ideal fidelity to the target states for our benchmark at the highest qubit number (20 qubits). We also provided simulations with noise and tests on real hardware. From the noise analysis, we expect VDSP will provide increasing performance as hardware noise levels improve due to its higher ideal fidelity. Although with small qubit numbers it is expected that the simplest method will perform the best, we have provided strong evidence that the VDSP method will excel at preparing highly entangled states even for larger qubit numbers when the gate noises of devices reach a sufficiently low level. However, some remarks about trainability of the parametrized quantum circuits are in order.

Firstly, the complexity of the singular value decomposition of an $m \times n, m\geq n$ matrix equals $O(mn^2)$. Calculating the full MPS representation of a state along with all the singular values thus scales exponentially in the number of qubits. However, it is also true that the amount of memory required to store all the amplitudes of a quantum state scales exponentially. That is, it takes $O(2^N)$ of memory to store all the amplitudes of an $N$-qubit state. It is therefore expected that all of the considered methods will run into memory issues when the number of qubits is increased. We used \texttt{quimb} library to train the PQCs in a tensor network format directly. In the tested qubit number range of 6--20, it was found that memory consumption was not yet a bottleneck. The VDSP method suffers from long training times, which is the primary downside of the method and provides a challenge for scaling the method beyond 20 qubits.

Finally, we note that there is the possibility to scale matrix product based methods to larger systems. For example, they can leverage tensor cross-interpolation methods \cite{bohun2025entanglementscalingmatrixproduct, manabe2025statepreparationmultivariatenormal} to avoid loading the state coefficients into memory. We can speculate that tensor cross-interpolation could also be advantageous for our method, if we can identify an efficiently computable entanglement measure as a loss function. Parametrized quantum circuits have been reportedly trained with up to 28 qubits \cite{hayes2023quantumstatepreparationgravitational} with parameter-shift rules and even beyond with gradient descent \cite{Melnikov_2023}. All of this suggests it is probably possible to significantly scale up our state preparation method. 

In the future we will look into running the VDSP algorithm on a HPC platform to scale the method beyond 20 qubits as well as to a larger class of states. Moreover, we will look into advanced training procedures to mitigate the long training time. Based on our initial analysis, the PQCs with entanglement entropy loss can be trained layer by layer, or in smaller chunks at a time, to obtain a better training time. This combined with some truncation at the MPS level to reduce memory consumption will allow us to scale the method up to comparable levels to basic PQC circuits on a high performance computational platform.

\section{Acknowledgments}

O.K. acknowledges financial support from KAUTE Foundation under a Post Docs in Companies grant.

V.L. acknowledges the support from the Research Council of Finland (Flagship of Advanced Mathematics for Sensing Imaging and Modelling grant 359183).

\bibliography{bibliography.bib}

\newpage

\onecolumngrid


\appendix

\section{MPO bond dimension}\label{appendix:mpo}

 We now study how the bond dimension of the matrix product operator (MPO) \cite{Schollwock2011TheDensity} representation of a PQC unitary depends on the number of layers. The PQC has parallel CX layers as presented in Fig. \ref{fig:PQC}.  In order to study this behavior, we fix the number of qubits and number of layers and randomly initialize a PQC with these dimensions. We then transform the corresponding randomly generated unitary matrix to MPO form with \texttt{quimb} and report the maximum bond dimension in Table \ref{tab:mpo-bond-dimension} for up to 10 qubits.

\begin{table}[ht]
\centering
{\setlength{\tabcolsep}{10pt}
\begin{tabular}{c | cccccccccc}
\toprule
$\chi$ & 2 & 4 & 8 & 16 & 32 & 64 & 128 & 256 & 512 & 1024 \\
\midrule
$N$ & 1 &  &  &  &  &  &  &  &  &  \\
$L$     & 1 &  &  &  &  &  &  &  &  &  \\ \hline
$N$ & 2 & 2 &  &  &  &  &  &  &  &  \\
$L$     & 1 & 2 &  &  &  &  &  &  &  &  \\ \hline
$N$ & 3 & 3 &  &  &  &  &  &  &  &  \\
$L$     & 1 & 2 &  &  &  &  &  &  &  &  \\ \hline
$N$ & 4 & 4 & 4 & 4 &  &  &  &  &  &  \\
$L$     & 1 & 2 & 3 & 4 &  &  &  &  &  &  \\ \hline
$N$ & 5 & 5 & 5 & 5 &  &  &  &  &  &  \\
$L$     & 1 & 2 & 3 & 4 &  &  &  &  &  &  \\ \hline
$N$ & 6 & 6 & 6 & 6 & 6 & 6 &  &  &  &  \\
$L$     & 1 & 2 & 3 & 4 & 5 & 6 &  &  &  &  \\ \hline
$N$ & 7 & 7 & 7 & 7 & 7 & 7 &  &  &  &  \\
$L$     & 1 & 2 & 3 & 4 & 5 & 6 &  &  &  &  \\ \hline
$N$ & 8 & 8 & 8 & 8 & 8 & 8 & 8 & 8 &  &  \\
$L$     & 1 & 2 & 3 & 4 & 5 & 6 & 7 & 8 &  &  \\ \hline
$N$ & 9 & 9 & 9 & 9 & 9 & 9 & 9 & 9 &  &  \\
$L$     & 1 & 2 & 3 & 4 & 5 & 6 & 7 & 8 &  &  \\ \hline
$N$ & 10 & 10 & 10 & 10 & 10 & 10 & 10 & 10 & 10 & 10 \\
$L$     & 1 & 2 & 3 & 4 & 5 & 6 & 7 & 8 & 9 & 10 \\ 
\bottomrule
\end{tabular}}
\caption{The maximum bond dimension of an MPO for a randomly generated PQC. $N$ denotes the number of qubits while $L$ denotes the number of variational PQC layers.}\label{tab:mpo-bond-dimension}
\end{table}

In Table \ref{tab:mpo-bond-dimension} we see interesting behavior for the bond dimension. Firstly, we see that the maximum bond dimension only grows when the number of qubits is even. The following line with odd number of qubits does not see an increase in bond dimension. This should only be interpreted as a property of the specific structure we chose for the ansatz circuit. Roughly we can interpreted the relationship of the bond dimension, number of qubits and number of variational layers in this way: for a single variational layer the bond dimension always equals 2, and the bond dimension doubles for each added variational layer before reaching a maximum depending on the number of qubits. The maximum bond dimension for $N$ qubits is $2^N$ when $N$ is even. 

We remark that the bond dimension reaching a maximum value doesn't mean that additional layers wouldn't be beneficial in practice. Indeed, there is no evidence that a linear number of layers would allow the PQC to reach an arbitrary unitary matrix. It merely means that a linear number of variational layers is enough to reach an MPO with maximal bond dimension. We take this as evidence that a linear number of layers could be an optimal starting point in optimization in most cases, and the number of layers can be adjusted either up or down when considering the training progress and achieved accuracy in terms of the loss function provided to the optimizer.

\section{Benchmarks}\label{appendix:benchmarks}

Let us demonstrate the training process of the VDSP method for a 2D normal distribution \begin{align}
    \Vec{\mu} = \begin{bmatrix}
    0.5 & 0.5
\end{bmatrix}, \quad \Sigma = \begin{bmatrix}
    0.1 & 0.01 \\ 0.01 & 0.1
\end{bmatrix}
\end{align} on 10 qubits. We first train a PQC with 2 layers and the loss function $S(\ket{\psi})$ of Eq. \eqref{eq:ee-loss-linear}. The state $\ket{\psi}$, visualized in Figure \ref{fig:vdsp-visualization} (a), initially has entanglement entropy of roughly $0.0055$. After training the PQC the entanglement entropy of the transformed state has reduced to $0.0014$, which is roughly a 75\% reduction. The transformed state with lower entanglement entropy is pictured in Figure \ref{fig:vdsp-visualization} (b). Observe that the transformed state has some clear structure, but it is far from a ``smooth'' distribution. Although MPS are typically used to prepare smooth distributions in the literature, the low entanglement entropy of the transformed state in Fig. \ref{fig:vdsp-visualization} (b) suggests that smoothness alone is not a good visual indicator of how MPS will perform in a particular QSP task.

Finally, to obtain the final state preparation circuit we form a single layer MPD with $\chi=2$ to prepare the transformed state in Fig. \ref{fig:vdsp-visualization} (b). The error in amplitudes of the resulting circuit is visualized in Fig. \ref{fig:vdsp-visualization} (c). The error of preparing the reduced entanglement entropy state is in Figure \ref{fig:vdsp-visualization} (d). We obtain an accuracy of  $0.9931$ for the VDSP circuit with  a single layer MPD with $\chi = 2$ followed by a 2 layer PQC. The infidelity equals $4.71 \times 10^{-5}$. The final transpiled circuit has depth 101 and uses 40 CX gates.

\input{figures/vdsp_visualization.tex}

For reference we can train a PQC with the same number of layers, and an extra layer of rotation gates so that the circuit does not end with CX gates, and the loss function \eqref{eq:distance-loss}. This PQC consists of 18 CX gates and has depth 26, but only achieves an accuracy of 0.9373 (infidelity $3.93 \times 10^{-3}$), which is significantly lower than the previous result where the PQC was used to minimize entanglement entropy. An MPD with 2 disentangling layers achieves an accuracy of 0.9848 (infidelity $2.31 \times 10^{-4}$) with 48 CX gates and depth 102. Exact state preparation uses 876 CX gates with depth 1696. So we conclude that minimizing entanglement entropy allows for excellent accuracy compared to PQC or MPD circuits while still only using a fraction of the gates compared to exact state preparation. We now collect benchmarks for a range of numbers layers and qubits for the 2D normal distribution and 2D Ricker wavelet states. In the tables the $S_\mathrm{init}$ column refers to the entanglement entropy of the target state, while the $S_\mathrm{final}$ refers to the entanglement entropy of the produced output state. In the column $S_\mathrm{red}$ we report how much the VDSP method was able to reduce entanglement entropy. We also collect training time for the parametrized circuits, final transpilation time along with peak memory consumption during training and transpilation. The column $L$ always refers to the number of parametrized layers for VDSP. We always include a single matrix product disentangling layer with $\chi = 2$ for VDSP.

From the tables we can see that all of the methods achieve good accuracy for the 2D normal distribution. This state has a relatively low entanglement entropy, so it is natural that this state is easy to prepare. We note that the VDSP method was unable to reduce the entanglement entropy with a single variational layer for the 2D normal distribution states, so the performance is identical to the MPD method in this case. However, for 2--5 layers the entanglement entropy is significantly reduced by the VDSP method and it achieves the highest level of typical accuracy, reaching a level of at least 0.99 for all numbers of qubits 6--20 and numbers of layers 2--5. 

The 2D Ricker wavelet has a significantly higher entanglement entropy compared to 2D normal distribution. The entanglement entropy reaches a value of 2.6259 for the 20-qubit state. For this state, we see that the VDSP is able to reduce the entanglement entropy for all numbers of layers 1--5 in all cases. Typically the entanglement entropy is reduced more when the number of layers increases. For qubit numbers 14--20 we see that the VDSP method is the only method that reaches infidelities below 0.01. Meanwhile, the PQC method struggles to minimize the distance to the target state. The infidelity gets stuck at 0.5 and is not able to improve for layers 1--5. It is unclear how many layers are needed to reach a better accuracy or if a different optimization approach improves the fidelity. The MPD method also scales very poorly in the number of layers. It is unclear if the MPD layers eventually reach a decent accuracy, and how many layers would be needed. The VDSP method is the only method suitable for NISQ devices for state preparation of 2D Ricker wavelet state with 14 or more qubits.

\begin{table}[ht]
\centering
\caption{Full benchmark results for $n_q = 6$ (2D Ricker wavelet).}
\label{tab:appendix-2d-ricker-6q}
\resizebox{\textwidth}{!}{%
\begin{tabular}{l r r r r r r r r r r r r r}
\toprule
Method & $L$ & Accuracy & Infidelity & Depth & CX & $S_\mathrm{init}$ & $S_\mathrm{red}$ & $S_\mathrm{final}$ & Train (s) & Mem (MB) & Tr.\ time (s) & Tr.\ mem (MB) \\
\midrule
VDSP & 1 & 0.9735 & $7.02 \times 10^{-4}$ & 60 & 19 & 0.4222 & 0.0123 & 0.4206 & 5.2 & 4.5 & 0.01 & 0.1 \\
 & 2 & 0.9848 & $2.32 \times 10^{-4}$ & 65 & 21 & 0.4222 & 0.0028 & 0.4254 & 8.4 & 5.8 & 0.01 & 0.1 \\
 & 3 & 0.9999 & $8.78 \times 10^{-9}$ & 77 & 28 & 0.4222 & 0.0000 & 0.4222 & 15.2 & 7.7 & 0.01 & 0.1 \\
 & 4 & 0.9999 & $1.20 \times 10^{-8}$ & 89 & 33 & 0.4222 & 0.0000 & 0.4221 & 15.9 & 8.3 & 0.01 & 0.1 \\
 & 5 & 0.9994 & $4.21 \times 10^{-7}$ & 99 & 37 & 0.4222 & 0.0000 & 0.4222 & 16.8 & 9.1 & 0.01 & 0.1 \\
\midrule
PQC & 1 & 0.4345 & $2.94 \times 10^{-1}$ & 16 & 5 & 0.4222 & -- & 0.0000 & 7.4 & 3.8 & 0.01 & 0.1 \\
 & 2 & 0.9722 & $7.75 \times 10^{-4}$ & 26 & 10 & 0.4222 & -- & 0.4169 & 9.7 & 4.7 & 0.01 & 0.1 \\
 & 3 & 0.9991 & $8.61 \times 10^{-7}$ & 36 & 15 & 0.4222 & -- & 0.4222 & 12.8 & 5.9 & 0.01 & 0.1 \\
 & 4 & 1.0000 & $0$ & 46 & 20 & 0.4222 & -- & 0.4222 & 2.5 & 6.2 & 0.01 & 0.1 \\
 & 5 & 0.9998 & $6.08 \times 10^{-8}$ & 56 & 25 & 0.4222 & -- & 0.4222 & 17.4 & 7.9 & 0.01 & 0.1 \\
\midrule
MPD & 1 & 0.5998 & $1.54 \times 10^{-1}$ & 50 & 13 & 0.4222 & -- & 0.0000 & -- & 0.1 & 0.01 & 0.1 \\
 & 2 & 0.7690 & $5.29 \times 10^{-2}$ & 73 & 25 & 0.4222 & -- & 0.2590 & -- & 0.1 & 0.01 & 0.1 \\
 & 3 & 0.8998 & $1.00 \times 10^{-2}$ & 97 & 43 & 0.4222 & -- & 0.3711 & -- & 0.2 & 0.01 & 0.1 \\
 & 4 & 0.8630 & $1.87 \times 10^{-2}$ & 114 & 54 & 0.4222 & -- & 0.3640 & -- & 0.2 & 0.01 & 0.1 \\
 & 5 & 0.8702 & $1.68 \times 10^{-2}$ & 135 & 62 & 0.4222 & -- & 0.3408 & -- & 0.2 & 0.01 & 0.1 \\
\midrule
Exact & -- & 1.0000 & $0$ & 116 & 44 & 0.4222 & -- & 0.4222 & -- & 0.1 & 0.01 & 0.1 \\
\bottomrule
\end{tabular}}
\end{table}

\begin{table}[ht]
\centering
\caption{Full benchmark results for $n_q = 8$ (2D Ricker wavelet).}
\label{tab:appendix-2d-ricker-8q}
\resizebox{\textwidth}{!}{%
\begin{tabular}{l r r r r r r r r r r r r r}
\toprule
Method & $L$ & Accuracy & Infidelity & Depth & CX & $S_\mathrm{init}$ & $S_\mathrm{red}$ & $S_\mathrm{final}$ & Train (s) & Mem (MB) & Tr.\ time (s) & Tr.\ mem (MB) \\
\midrule
VDSP & 1 & 0.9082 & $8.41 \times 10^{-3}$ & 76 & 24 & 1.4069 & 0.0719 & 1.4160 & 7.8 & 5.8 & 0.01 & 0.1 \\
 & 2 & 0.9183 & $6.66 \times 10^{-3}$ & 88 & 32 & 1.4069 & 0.0496 & 1.4056 & 19.2 & 8.9 & 0.01 & 0.1 \\
 & 3 & 0.9988 & $1.55 \times 10^{-6}$ & 94 & 38 & 1.4069 & 0.0000 & 1.4069 & 31.2 & 10.7 & 0.01 & 0.1 \\
 & 4 & 0.9998 & $2.59 \times 10^{-8}$ & 106 & 46 & 1.4069 & 0.0000 & 1.4069 & 30.2 & 11.2 & 0.01 & 0.1 \\
 & 5 & 0.9997 & $7.56 \times 10^{-8}$ & 116 & 51 & 1.4069 & 0.0000 & 1.4069 & 33.0 & 12.5 & 0.01 & 0.1 \\
\midrule
PQC & 1 & 0.1511 & $5.91 \times 10^{-1}$ & 16 & 7 & 1.4069 & -- & 0.0000 & 9.1 & 4.3 & 0.01 & 0.1 \\
 & 2 & 0.4089 & $3.19 \times 10^{-1}$ & 26 & 14 & 1.4069 & -- & 0.8768 & 10.8 & 5.7 & 0.01 & 0.1 \\
 & 3 & 0.9157 & $7.10 \times 10^{-3}$ & 36 & 21 & 1.4069 & -- & 1.4073 & 15.5 & 6.9 & 0.01 & 0.1 \\
 & 4 & 0.9820 & $3.26 \times 10^{-4}$ & 46 & 28 & 1.4069 & -- & 1.4058 & 19.3 & 8.4 & 0.01 & 0.1 \\
 & 5 & 0.9996 & $1.33 \times 10^{-7}$ & 56 & 35 & 1.4069 & -- & 1.4069 & 97.6 & 17.6 & 0.01 & 0.1 \\
\midrule
MPD & 1 & 0.3511 & $3.77 \times 10^{-1}$ & 66 & 18 & 1.4069 & -- & 0.0000 & -- & 0.1 & 0.01 & 0.1 \\
 & 2 & 0.5077 & $2.28 \times 10^{-1}$ & 91 & 37 & 1.4069 & -- & 0.6752 & -- & 0.2 & 0.01 & 0.1 \\
 & 3 & 0.7169 & $7.91 \times 10^{-2}$ & 111 & 54 & 1.4069 & -- & 1.5264 & -- & 0.2 & 0.01 & 0.1 \\
 & 4 & 0.6481 & $1.21 \times 10^{-1}$ & 131 & 71 & 1.4069 & -- & 1.1439 & -- & 0.2 & 0.01 & 0.1 \\
 & 5 & 0.7554 & $5.94 \times 10^{-2}$ & 149 & 86 & 1.4069 & -- & 1.4879 & -- & 0.3 & 0.01 & 0.1 \\
\midrule
Exact & -- & 1.0000 & $0$ & 399 & 198 & 1.4069 & -- & 1.4069 & -- & 0.1 & 0.01 & 0.1 \\
\bottomrule
\end{tabular}}
\end{table}

\begin{table}[ht]
\centering
\caption{Full benchmark results for $n_q = 10$ (2D Ricker wavelet).}
\label{tab:appendix-2d-ricker-10q}
\resizebox{\textwidth}{!}{%
\begin{tabular}{l r r r r r r r r r r r r r}
\toprule
Method & $L$ & Accuracy & Infidelity & Depth & CX & $S_\mathrm{init}$ & $S_\mathrm{red}$ & $S_\mathrm{final}$ & Train (s) & Mem (MB) & Tr.\ time (s) & Tr.\ mem (MB) \\
\midrule
VDSP & 1 & 0.5274 & $2.11 \times 10^{-1}$ & 93 & 33 & 2.1789 & 1.1994 & 1.6538 & 11.5 & 7.2 & 0.01 & 0.1 \\
 & 2 & 0.8945 & $1.11 \times 10^{-2}$ & 101 & 38 & 2.1789 & 0.1180 & 2.0675 & 21.4 & 9.6 & 0.01 & 0.1 \\
 & 3 & 0.9745 & $6.51 \times 10^{-4}$ & 110 & 49 & 2.1789 & 0.0146 & 2.1705 & 50.7 & 13.6 & 0.02 & 0.1 \\
 & 4 & 0.9895 & $1.10 \times 10^{-4}$ & 123 & 60 & 2.1789 & 0.0038 & 2.1771 & 40.5 & 13.7 & 0.01 & 0.1 \\
 & 5 & 0.9898 & $1.04 \times 10^{-4}$ & 135 & 69 & 2.1789 & 0.0038 & 2.1768 & 61.2 & 16.4 & 0.01 & 0.1 \\
\midrule
PQC & 1 & 0.1258 & $6.18 \times 10^{-1}$ & 16 & 9 & 2.1789 & -- & 0.0000 & 9.3 & 4.8 & 0.01 & 0.1 \\
 & 2 & 0.4796 & $2.53 \times 10^{-1}$ & 26 & 18 & 2.1789 & -- & 1.4088 & 12.7 & 6.5 & 0.01 & 0.1 \\
 & 3 & 0.7091 & $8.28 \times 10^{-2}$ & 36 & 27 & 2.1789 & -- & 2.0935 & 44.7 & 11.4 & 0.01 & 0.1 \\
 & 4 & 0.9591 & $1.67 \times 10^{-3}$ & 46 & 36 & 2.1789 & -- & 2.1592 & 29.9 & 10.9 & 0.01 & 0.1 \\
 & 5 & 0.9780 & $4.86 \times 10^{-4}$ & 56 & 45 & 2.1789 & -- & 2.1739 & 32.0 & 13.0 & 0.01 & 0.1 \\
\midrule
MPD & 1 & 0.2262 & $5.09 \times 10^{-1}$ & 85 & 25 & 2.1789 & -- & 0.0000 & -- & 0.4 & 0.01 & 0.1 \\
 & 2 & 0.5001 & $2.39 \times 10^{-1}$ & 106 & 47 & 2.1789 & -- & 1.6788 & -- & 0.4 & 0.01 & 0.1 \\
 & 3 & 0.5531 & $1.90 \times 10^{-1}$ & 129 & 69 & 2.1789 & -- & 2.2115 & -- & 0.4 & 0.01 & 0.1 \\
 & 4 & 0.6867 & $9.68 \times 10^{-2}$ & 149 & 96 & 2.1789 & -- & 2.2156 & -- & 0.4 & 0.02 & 0.1 \\
 & 5 & 0.5631 & $1.84 \times 10^{-1}$ & 172 & 119 & 2.1789 & -- & 2.2064 & -- & 0.5 & 0.01 & 0.1 \\
\midrule
Exact & -- & 1.0000 & $0$ & 1680 & 873 & 2.1789 & -- & 2.1789 & -- & 0.2 & 0.02 & 0.1 \\
\bottomrule
\end{tabular}}
\end{table}

\begin{table}[ht]
\centering
\caption{Full benchmark results for $n_q = 12$ (2D Ricker wavelet).}
\label{tab:appendix-2d-ricker-12q}
\resizebox{\textwidth}{!}{%
\begin{tabular}{l r r r r r r r r r r r r r}
\toprule
Method & $L$ & Accuracy & Infidelity & Depth & CX & $S_\mathrm{init}$ & $S_\mathrm{red}$ & $S_\mathrm{final}$ & Train (s) & Mem (MB) & Tr.\ time (s) & Tr.\ mem (MB) \\
\midrule
VDSP & 1 & 0.5865 & $1.64 \times 10^{-1}$ & 109 & 40 & 2.4625 & 0.9775 & 1.9597 & 27.4 & 9.9 & 0.01 & 0.1 \\
 & 2 & 0.8790 & $1.46 \times 10^{-2}$ & 116 & 47 & 2.4625 & 0.1621 & 2.3088 & 29.8 & 11.6 & 0.01 & 0.1 \\
 & 3 & 0.9632 & $1.35 \times 10^{-3}$ & 130 & 63 & 2.4625 & 0.0321 & 2.4601 & 37.4 & 13.7 & 0.01 & 0.1 \\
 & 4 & 0.9739 & $6.84 \times 10^{-4}$ & 136 & 73 & 2.4625 & 0.0175 & 2.4548 & 74.0 & 17.2 & 0.02 & 0.1 \\
 & 5 & 0.9880 & $1.44 \times 10^{-4}$ & 150 & 82 & 2.4625 & 0.0050 & 2.4601 & 90.4 & 19.1 & 0.02 & 0.6 \\
\midrule
PQC & 1 & 0.0464 & $7.03 \times 10^{-1}$ & 16 & 11 & 2.4625 & -- & 0.0000 & 13.5 & 6.1 & 0.01 & 0.1 \\
 & 2 & 0.2294 & $5.06 \times 10^{-1}$ & 26 & 22 & 2.4625 & -- & 0.0605 & 19.8 & 8.2 & 0.01 & 0.1 \\
 & 3 & 0.2345 & $5.00 \times 10^{-1}$ & 36 & 33 & 2.4625 & -- & 0.0276 & 26.5 & 10.2 & 0.01 & 0.1 \\
 & 4 & 0.8132 & $3.46 \times 10^{-2}$ & 46 & 44 & 2.4625 & -- & 2.4876 & 34.3 & 13.0 & 0.01 & 0.1 \\
 & 5 & 0.9452 & $3.00 \times 10^{-3}$ & 56 & 55 & 2.4625 & -- & 2.4202 & 40.0 & 15.6 & 0.01 & 0.1 \\
\midrule
MPD & 1 & 0.2113 & $5.25 \times 10^{-1}$ & 99 & 28 & 2.4625 & -- & 0.0000 & -- & 1.4 & 0.01 & 0.1 \\
 & 2 & 0.3248 & $4.09 \times 10^{-1}$ & 125 & 58 & 2.4625 & -- & 1.2359 & -- & 1.4 & 0.01 & 0.1 \\
 & 3 & 0.4950 & $2.47 \times 10^{-1}$ & 145 & 89 & 2.4625 & -- & 2.3637 & -- & 1.4 & 0.01 & 0.1 \\
 & 4 & 0.5152 & $2.24 \times 10^{-1}$ & 167 & 111 & 2.4625 & -- & 2.8944 & -- & 1.4 & 0.02 & 0.1 \\
 & 5 & 0.5563 & $1.90 \times 10^{-1}$ & 190 & 141 & 2.4625 & -- & 3.1779 & -- & 1.4 & 0.01 & 0.1 \\
\midrule
Exact & -- & 1.0000 & $0$ & 6695 & 3612 & 2.4625 & -- & 2.4625 & -- & 0.5 & 0.04 & 0.4 \\
\bottomrule
\end{tabular}}
\end{table}

\begin{table}[ht]
\centering
\caption{Full benchmark results for $n_q = 14$ (2D Ricker wavelet).}
\label{tab:appendix-2d-ricker-14q}
\resizebox{\textwidth}{!}{%
\begin{tabular}{l r r r r r r r r r r r r r}
\toprule
Method & $L$ & Accuracy & Infidelity & Depth & CX & $S_\mathrm{init}$ & $S_\mathrm{red}$ & $S_\mathrm{final}$ & Train (s) & Mem (MB) & Tr.\ time (s) & Tr.\ mem (MB) \\
\midrule
VDSP & 1 & 0.5118 & $2.24 \times 10^{-1}$ & 120 & 43 & 2.5684 & 1.4755 & 2.1275 & 52.7 & 19.1 & 0.01 & 0.1 \\
 & 2 & 0.8429 & $2.45 \times 10^{-2}$ & 137 & 60 & 2.5684 & 0.2750 & 2.4102 & 64.5 & 15.1 & 0.01 & 0.1 \\
 & 3 & 0.9248 & $5.65 \times 10^{-3}$ & 147 & 74 & 2.5684 & 0.1604 & 2.4754 & 104.7 & 18.1 & 0.01 & 0.1 \\
 & 4 & 0.9701 & $8.94 \times 10^{-4}$ & 155 & 87 & 2.5684 & 0.0298 & 2.5539 & 292.9 & 26.8 & 0.02 & 0.1 \\
 & 5 & 0.9739 & $6.82 \times 10^{-4}$ & 159 & 95 & 2.5684 & 0.0228 & 2.5522 & 153.3 & 23.1 & 0.02 & 0.1 \\
\midrule
PQC & 1 & 0.1709 & $5.69 \times 10^{-1}$ & 16 & 13 & 2.5684 & -- & 0.0000 & 16.6 & 6.8 & 0.01 & 0.1 \\
 & 2 & 0.2303 & $5.05 \times 10^{-1}$ & 26 & 26 & 2.5684 & -- & 0.0841 & 22.5 & 9.1 & 0.01 & 0.1 \\
 & 3 & 0.2341 & $5.01 \times 10^{-1}$ & 36 & 39 & 2.5684 & -- & 0.0959 & 26.8 & 12.2 & 0.01 & 0.1 \\
 & 4 & 0.2345 & $5.00 \times 10^{-1}$ & 46 & 52 & 2.5684 & -- & 0.0356 & 34.5 & 15.2 & 0.01 & 0.1 \\
 & 5 & 0.2346 & $5.00 \times 10^{-1}$ & 56 & 65 & 2.5684 & -- & 0.0990 & 44.0 & 18.4 & 0.01 & 0.1 \\
\midrule
MPD & 1 & 0.2092 & $5.28 \times 10^{-1}$ & 113 & 31 & 2.5684 & -- & 0.0000 & -- & 5.6 & 0.01 & 0.1 \\
 & 2 & 0.3903 & $3.39 \times 10^{-1}$ & 137 & 64 & 2.5684 & -- & 1.3935 & -- & 5.6 & 0.01 & 0.1 \\
 & 3 & 0.5110 & $2.27 \times 10^{-1}$ & 159 & 101 & 2.5684 & -- & 3.2788 & -- & 5.6 & 0.01 & 0.1 \\
 & 4 & 0.5829 & $1.70 \times 10^{-1}$ & 179 & 131 & 2.5684 & -- & 2.7403 & -- & 5.6 & 0.02 & 0.1 \\
 & 5 & 0.6196 & $1.41 \times 10^{-1}$ & 205 & 168 & 2.5684 & -- & 3.9943 & -- & 5.6 & 0.02 & 0.1 \\
\midrule
Exact & -- & 1.0000 & $0$ & 27056 & 14762 & 2.5684 & -- & 2.5684 & -- & 1.8 & 0.18 & 1.5 \\
\bottomrule
\end{tabular}}
\end{table}

\begin{table}[ht]
\centering
\caption{Full benchmark results for $n_q = 16$ (2D Ricker wavelet).}
\label{tab:appendix-2d-ricker-16q}
\resizebox{\textwidth}{!}{%
\begin{tabular}{l r r r r r r r r r r r r r}
\toprule
Method & $L$ & Accuracy & Infidelity & Depth & CX & $S_\mathrm{init}$ & $S_\mathrm{red}$ & $S_\mathrm{final}$ & Train (s) & Mem (MB) & Tr.\ time (s) & Tr.\ mem (MB) \\
\midrule
VDSP & 1 & 0.6003 & $1.53 \times 10^{-1}$ & 148 & 53 & 2.6070 & 1.2450 & 1.9798 & 133.8 & 21.8 & 0.01 & 0.1 \\
 & 2 & 0.4949 & $2.39 \times 10^{-1}$ & 150 & 67 & 2.6070 & 0.8995 & 1.6174 & 163.1 & 22.5 & 0.01 & 0.1 \\
 & 3 & 0.8993 & $1.01 \times 10^{-2}$ & 157 & 83 & 2.6070 & 0.2595 & 2.4596 & 232.3 & 25.8 & 0.01 & 0.1 \\
 & 4 & 0.9254 & $5.56 \times 10^{-3}$ & 173 & 101 & 2.6070 & 0.0801 & 2.5298 & 478.5 & 32.1 & 0.01 & 0.1 \\
 & 5 & 0.9460 & $2.91 \times 10^{-3}$ & 179 & 113 & 2.6070 & 0.0407 & 2.5617 & 395.3 & 32.9 & 0.02 & 0.1 \\
\midrule
PQC & 1 & -0.0008 & $7.51 \times 10^{-1}$ & 16 & 15 & 2.6070 & -- & 0.0000 & 19.8 & 8.1 & 0.01 & 0.1 \\
 & 2 & 0.2276 & $5.08 \times 10^{-1}$ & 26 & 30 & 2.6070 & -- & 0.0546 & 26.6 & 10.9 & 0.01 & 0.1 \\
 & 3 & 0.2333 & $5.01 \times 10^{-1}$ & 36 & 45 & 2.6070 & -- & 0.1016 & 32.9 & 14.6 & 0.01 & 0.1 \\
 & 4 & 0.2340 & $5.01 \times 10^{-1}$ & 46 & 60 & 2.6070 & -- & 0.1187 & 42.5 & 18.1 & 0.01 & 0.1 \\
 & 5 & 0.2344 & $5.00 \times 10^{-1}$ & 56 & 75 & 2.6070 & -- & 0.1145 & 55.2 & 21.8 & 0.01 & 0.1 \\
\midrule
MPD & 1 & 0.2087 & $5.28 \times 10^{-1}$ & 135 & 37 & 2.6070 & -- & 0.0000 & -- & 11.4 & 0.01 & 0.1 \\
 & 2 & 0.4212 & $3.14 \times 10^{-1}$ & 153 & 74 & 2.6070 & -- & 1.8936 & -- & 11.9 & 0.01 & 0.1 \\
 & 3 & 0.5345 & $2.06 \times 10^{-1}$ & 180 & 114 & 2.6070 & -- & 3.3826 & -- & 11.9 & 0.02 & 0.1 \\
 & 4 & 0.5931 & $1.61 \times 10^{-1}$ & 196 & 146 & 2.6070 & -- & 4.1616 & -- & 12.0 & 0.02 & 0.1 \\
 & 5 & 0.6868 & $9.58 \times 10^{-2}$ & 214 & 191 & 2.6070 & -- & 3.1806 & -- & 12.0 & 0.02 & 0.1 \\
\midrule
Exact & -- & 1.0000 & $0$ & 108159 & 59517 & 2.6070 & -- & 2.6070 & -- & 7.1 & 0.94 & 6.1 \\
\bottomrule
\end{tabular}}
\end{table}

\begin{table}[ht]
\centering
\caption{Full benchmark results for $n_q = 18$ (2D Ricker wavelet).}
\label{tab:appendix-2d-ricker-18q}
\resizebox{\textwidth}{!}{%
\begin{tabular}{l r r r r r r r r r r r r r}
\toprule
Method & $L$ & Accuracy & Infidelity & Depth & CX & $S_\mathrm{init}$ & $S_\mathrm{red}$ & $S_\mathrm{final}$ & Train (s) & Mem (MB) & Tr.\ time (s) & Tr.\ mem (MB) \\
\midrule
VDSP & 1 & 0.5131 & $2.23 \times 10^{-1}$ & 160 & 61 & 2.6209 & 1.5330 & 2.1157 & 261.9 & 43.7 & 0.02 & 0.1 \\
 & 2 & 0.8602 & $1.95 \times 10^{-2}$ & 169 & 78 & 2.6209 & 0.3332 & 2.3688 & 439.6 & 48.2 & 0.02 & 0.1 \\
 & 3 & 0.8997 & $1.00 \times 10^{-2}$ & 173 & 94 & 2.6209 & 0.2393 & 2.4612 & 817.8 & 53.4 & 0.02 & 0.1 \\
 & 4 & 0.9424 & $3.32 \times 10^{-3}$ & 187 & 111 & 2.6209 & 0.0700 & 2.5641 & 966.6 & 55.8 & 0.02 & 0.1 \\
 & 5 & 0.9213 & $6.19 \times 10^{-3}$ & 195 & 125 & 2.6209 & 0.1308 & 2.5043 & 1074.8 & 57.8 & 0.02 & 0.1 \\
\midrule
PQC & 1 & 0.1700 & $5.70 \times 10^{-1}$ & 16 & 17 & 2.6209 & -- & 0.0000 & 33.5 & 12.9 & 0.01 & 0.1 \\
 & 2 & 0.2280 & $5.07 \times 10^{-1}$ & 26 & 34 & 2.6209 & -- & 0.0736 & 38.1 & 15.7 & 0.01 & 0.1 \\
 & 3 & 0.2340 & $5.01 \times 10^{-1}$ & 36 & 51 & 2.6209 & -- & 0.0549 & 48.3 & 19.2 & 0.02 & 0.1 \\
 & 4 & 0.2343 & $5.00 \times 10^{-1}$ & 46 & 68 & 2.6209 & -- & 0.1109 & 57.2 & 23.3 & 0.01 & 0.1 \\
 & 5 & 0.2345 & $5.00 \times 10^{-1}$ & 56 & 85 & 2.6209 & -- & 0.0927 & 79.2 & 27.7 & 0.01 & 0.1 \\
\midrule
MPD & 1 & 0.2086 & $5.28 \times 10^{-1}$ & 155 & 42 & 2.6209 & -- & 0.0000 & -- & 30.0 & 0.01 & 0.1 \\
 & 2 & 0.3328 & $3.96 \times 10^{-1}$ & 174 & 85 & 2.6209 & -- & 1.6413 & -- & 30.0 & 0.02 & 0.1 \\
 & 3 & 0.4692 & $2.64 \times 10^{-1}$ & 195 & 123 & 2.6209 & -- & 3.4053 & -- & 30.1 & 0.02 & 0.1 \\
 & 4 & 0.5930 & $1.62 \times 10^{-1}$ & 217 & 174 & 2.6209 & -- & 3.7886 & -- & 30.1 & 0.02 & 0.1 \\
 & 5 & 0.6111 & $1.47 \times 10^{-1}$ & 239 & 215 & 2.6209 & -- & 3.9157 & -- & 30.2 & 0.02 & 0.1 \\
\midrule
Exact & -- & 1.0000 & $0$ & 433771 & 239305 & 2.6209 & -- & 2.6209 & -- & 28.1 & 5.42 & 24.1 \\
\bottomrule
\end{tabular}}
\end{table}

\begin{table}[ht]
\centering
\caption{Full benchmark results for $n_q = 20$ (2D Ricker wavelet).}
\label{tab:appendix-2d-ricker-20q}
\resizebox{\textwidth}{!}{%
\begin{tabular}{l r r r r r r r r r r r r r}
\toprule
Method & $L$ & Accuracy & Infidelity & Depth & CX & $S_\mathrm{init}$ & $S_\mathrm{red}$ & $S_\mathrm{final}$ & Train (s) & Mem (MB) & Tr.\ time (s) & Tr.\ mem (MB) \\
\midrule
VDSP & 1 & 0.6005 & $1.53 \times 10^{-1}$ & 175 & 69 & 2.6259 & 1.2684 & 1.9784 & 1664.6 & 143.3 & 0.01 & 0.1 \\
 & 2 & 0.8502 & $2.23 \times 10^{-2}$ & 191 & 87 & 2.6259 & 0.3233 & 2.3539 & 7044.6 & 147.7 & 0.02 & 0.1 \\
 & 3 & 0.9138 & $7.41 \times 10^{-3}$ & 194 & 106 & 2.6259 & 0.1448 & 2.5444 & 1764.3 & 150.3 & 0.02 & 0.1 \\
 & 4 & 0.9485 & $2.65 \times 10^{-3}$ & 202 & 122 & 2.6259 & 0.0903 & 2.5970 & 2071.9 & 154.1 & 0.02 & 0.1 \\
 & 5 & 0.9687 & $9.77 \times 10^{-4}$ & 211 & 141 & 2.6259 & 0.0423 & 2.5951 & 3732.5 & 157.9 & 0.02 & 0.1 \\
\midrule
PQC & 1 & -0.0025 & $7.52 \times 10^{-1}$ & 16 & 19 & 2.6259 & -- & 0.0000 & 55.0 & 37.4 & 0.01 & 0.1 \\
 & 2 & 0.2274 & $5.08 \times 10^{-1}$ & 26 & 38 & 2.6259 & -- & 0.0552 & 121.6 & 40.5 & 0.01 & 0.1 \\
 & 3 & 0.2332 & $5.02 \times 10^{-1}$ & 36 & 57 & 2.6259 & -- & 0.1039 & 91.4 & 43.6 & 0.01 & 0.1 \\
 & 4 & 0.2343 & $5.00 \times 10^{-1}$ & 46 & 76 & 2.6259 & -- & 0.1126 & 106.5 & 46.9 & 0.01 & 0.1 \\
 & 5 & 0.2345 & $5.00 \times 10^{-1}$ & 56 & 95 & 2.6259 & -- & 0.1146 & 181.6 & 50.1 & 0.01 & 0.1 \\
\midrule
MPD & 1 & 0.2086 & $5.28 \times 10^{-1}$ & 166 & 48 & 2.6259 & -- & 0.0000 & -- & 93.7 & 0.01 & 0.1 \\
 & 2 & 0.4443 & $2.90 \times 10^{-1}$ & 190 & 100 & 2.6259 & -- & 2.4837 & -- & 93.7 & 0.02 & 0.1 \\
 & 3 & 0.3857 & $3.44 \times 10^{-1}$ & 214 & 148 & 2.6259 & -- & 5.2421 & -- & 93.9 & 0.02 & 0.1 \\
 & 4 & 0.5556 & $1.91 \times 10^{-1}$ & 238 & 185 & 2.6259 & -- & 4.5767 & -- & 94.0 & 0.02 & 0.1 \\
 & 5 & 0.4914 & $2.45 \times 10^{-1}$ & 252 & 250 & 2.6259 & -- & 4.5468 & -- & 94.0 & 0.02 & 0.1 \\
\midrule
Exact & -- & 1.0000 & $0$ & 1735318 & 959010 & 2.6259 & -- & 2.6259 & -- & 112.1 & 35.57 & 96.1 \\
\bottomrule
\end{tabular}}
\end{table}


\begin{table}[ht]
\centering
\caption{Full benchmark results for $n_q = 6$ (2D normal).}
\label{tab:appendix-2d-normal-6q}
\resizebox{\textwidth}{!}{%
\begin{tabular}{l r r r r r r r r r r r r r}
\toprule
Method & $L$ & Accuracy & Infidelity & Depth & CX & $S_\mathrm{init}$ & $S_\mathrm{red}$ & $S_\mathrm{final}$ & Train (s) & Mem (MB) & Tr.\ time (s) & Tr.\ mem (MB) \\
\midrule
VDSP & 1 & 0.9903 & $9.42 \times 10^{-5}$ & 51 & 13 & 0.0014 & 0.0014 & 0.0000 & 10.7 & 12.5 & 0.01 & 0.1 \\
 & 2 & 0.9987 & $1.81 \times 10^{-6}$ & 67 & 23 & 0.0014 & 0.0001 & 0.0014 & 9.8 & 6.2 & 0.01 & 0.1 \\
 & 3 & 0.9988 & $1.54 \times 10^{-6}$ & 82 & 28 & 0.0014 & 0.0001 & 0.0014 & 12.3 & 7.2 & 0.01 & 0.1 \\
 & 4 & 0.9986 & $1.89 \times 10^{-6}$ & 89 & 31 & 0.0014 & 0.0001 & 0.0014 & 14.7 & 8.4 & 0.01 & 0.1 \\
 & 5 & 0.9996 & $1.37 \times 10^{-7}$ & 100 & 38 & 0.0014 & 0.0000 & 0.0014 & 19.1 & 9.4 & 0.01 & 0.1 \\
\midrule
PQC & 1 & 0.8161 & $3.35 \times 10^{-2}$ & 16 & 5 & 0.0014 & -- & 0.0000 & 10.1 & 4.4 & 0.01 & 0.1 \\
 & 2 & 0.9770 & $5.28 \times 10^{-4}$ & 26 & 10 & 0.0014 & -- & 0.0029 & 13.3 & 5.5 & 0.01 & 0.1 \\
 & 3 & 0.9987 & $1.75 \times 10^{-6}$ & 36 & 15 & 0.0014 & -- & 0.0014 & 14.8 & 6.5 & 0.01 & 0.1 \\
 & 4 & 0.9987 & $1.63 \times 10^{-6}$ & 46 & 20 & 0.0014 & -- & 0.0014 & 16.1 & 7.1 & 0.01 & 0.1 \\
 & 5 & 0.9997 & $1.02 \times 10^{-7}$ & 56 & 25 & 0.0014 & -- & 0.0014 & 20.0 & 8.3 & 0.01 & 0.1 \\
\midrule
MPD & 1 & 0.9903 & $9.42 \times 10^{-5}$ & 51 & 13 & 0.0014 & -- & 0.0000 & -- & 0.1 & 0.01 & 0.1 \\
 & 2 & 0.9915 & $7.20 \times 10^{-5}$ & 73 & 27 & 0.0014 & -- & 0.0004 & -- & 0.1 & 0.01 & 0.1 \\
 & 3 & 0.9942 & $3.37 \times 10^{-5}$ & 93 & 40 & 0.0014 & -- & 0.0012 & -- & 0.2 & 0.01 & 0.1 \\
 & 4 & 0.9963 & $1.39 \times 10^{-5}$ & 114 & 54 & 0.0014 & -- & 0.0012 & -- & 0.2 & 0.01 & 0.1 \\
 & 5 & 0.9958 & $1.77 \times 10^{-5}$ & 138 & 65 & 0.0014 & -- & 0.0012 & -- & 0.2 & 0.01 & 0.1 \\
\midrule
Exact & -- & 1.0000 & $0$ & 122 & 45 & 0.0014 & -- & 0.0014 & -- & 0.1 & 0.01 & 0.1 \\
\bottomrule
\end{tabular}}
\end{table}

\begin{table}[ht]
\centering
\caption{Full benchmark results for $n_q = 8$ (2D normal).}
\label{tab:appendix-2d-normal-8q}
\resizebox{\textwidth}{!}{%
\begin{tabular}{l r r r r r r r r r r r r r}
\toprule
Method & $L$ & Accuracy & Infidelity & Depth & CX & $S_\mathrm{init}$ & $S_\mathrm{red}$ & $S_\mathrm{final}$ & Train (s) & Mem (MB) & Tr.\ time (s) & Tr.\ mem (MB) \\
\midrule
VDSP & 1 & 0.9822 & $3.16 \times 10^{-4}$ & 67 & 17 & 0.0044 & 0.0044 & 0.0000 & 11.9 & 15.1 & 0.01 & 0.1 \\
 & 2 & 0.9964 & $1.30 \times 10^{-5}$ & 83 & 31 & 0.0044 & 0.0004 & 0.0040 & 14.1 & 7.9 & 0.01 & 0.1 \\
 & 3 & 0.9980 & $4.11 \times 10^{-6}$ & 98 & 39 & 0.0044 & 0.0002 & 0.0043 & 21.4 & 9.7 & 0.01 & 0.1 \\
 & 4 & 0.9981 & $3.49 \times 10^{-6}$ & 108 & 45 & 0.0044 & 0.0002 & 0.0044 & 28.6 & 11.3 & 0.01 & 0.1 \\
 & 5 & 0.9993 & $4.40 \times 10^{-7}$ & 116 & 54 & 0.0044 & 0.0000 & 0.0044 & 29.8 & 12.8 & 0.01 & 0.1 \\
\midrule
PQC & 1 & 0.7936 & $4.22 \times 10^{-2}$ & 16 & 7 & 0.0044 & -- & 0.0000 & 10.7 & 5.0 & 0.01 & 0.1 \\
 & 2 & 0.9406 & $3.53 \times 10^{-3}$ & 26 & 14 & 0.0044 & -- & 0.0058 & 14.6 & 6.4 & 0.01 & 0.1 \\
 & 3 & 0.9732 & $7.20 \times 10^{-4}$ & 36 & 21 & 0.0044 & -- & 0.0038 & 18.5 & 7.7 & 0.01 & 0.1 \\
 & 4 & 0.9927 & $5.35 \times 10^{-5}$ & 46 & 28 & 0.0044 & -- & 0.0050 & 21.3 & 9.0 & 0.01 & 0.1 \\
 & 5 & 0.9985 & $2.26 \times 10^{-6}$ & 56 & 35 & 0.0044 & -- & 0.0043 & 22.9 & 10.1 & 0.01 & 0.1 \\
\midrule
MPD & 1 & 0.9822 & $3.16 \times 10^{-4}$ & 67 & 18 & 0.0044 & -- & 0.0000 & -- & 0.1 & 0.01 & 0.1 \\
 & 2 & 0.9845 & $2.40 \times 10^{-4}$ & 88 & 38 & 0.0044 & -- & 0.0017 & -- & 0.2 & 0.01 & 0.1 \\
 & 3 & 0.9899 & $1.02 \times 10^{-4}$ & 107 & 57 & 0.0044 & -- & 0.0032 & -- & 0.2 & 0.01 & 0.1 \\
 & 4 & 0.9899 & $1.02 \times 10^{-4}$ & 131 & 70 & 0.0044 & -- & 0.0036 & -- & 0.2 & 0.01 & 0.1 \\
 & 5 & 0.9910 & $8.06 \times 10^{-5}$ & 151 & 91 & 0.0044 & -- & 0.0042 & -- & 0.3 & 0.01 & 0.1 \\
\midrule
Exact & -- & 1.0000 & $0$ & 409 & 200 & 0.0044 & -- & 0.0044 & -- & 0.1 & 0.01 & 0.1 \\
\bottomrule
\end{tabular}}
\end{table}

\begin{table}[ht]
\centering
\caption{Full benchmark results for $n_q = 10$ (2D normal).}
\label{tab:appendix-2d-normal-10q}
\resizebox{\textwidth}{!}{%
\begin{tabular}{l r r r r r r r r r r r r r}
\toprule
Method & $L$ & Accuracy & Infidelity & Depth & CX & $S_\mathrm{init}$ & $S_\mathrm{red}$ & $S_\mathrm{final}$ & Train (s) & Mem (MB) & Tr.\ time (s) & Tr.\ mem (MB) \\
\midrule
VDSP & 1 & 0.9804 & $3.83 \times 10^{-4}$ & 83 & 22 & 0.0055 & 0.0055 & 0.0000 & 14.4 & 17.4 & 0.01 & 0.1 \\
 & 2 & 0.9931 & $4.71 \times 10^{-5}$ & 101 & 40 & 0.0055 & 0.0014 & 0.0042 & 18.3 & 9.6 & 0.01 & 0.1 \\
 & 3 & 0.9930 & $4.85 \times 10^{-5}$ & 111 & 51 & 0.0055 & 0.0015 & 0.0061 & 29.1 & 11.3 & 0.01 & 0.1 \\
 & 4 & 0.9955 & $2.02 \times 10^{-5}$ & 126 & 62 & 0.0055 & 0.0007 & 0.0054 & 39.3 & 13.8 & 0.01 & 0.1 \\
 & 5 & 0.9978 & $4.91 \times 10^{-6}$ & 134 & 67 & 0.0055 & 0.0003 & 0.0053 & 39.2 & 17.6 & 0.02 & 0.7 \\
\midrule
PQC & 1 & 0.7888 & $4.41 \times 10^{-2}$ & 16 & 9 & 0.0055 & -- & 0.0000 & 12.6 & 5.6 & 0.01 & 0.1 \\
 & 2 & 0.9373 & $3.93 \times 10^{-3}$ & 26 & 18 & 0.0055 & -- & 0.0042 & 17.1 & 7.2 & 0.01 & 0.1 \\
 & 3 & 0.9750 & $6.26 \times 10^{-4}$ & 36 & 27 & 0.0055 & -- & 0.0034 & 20.3 & 8.6 & 0.01 & 0.1 \\
 & 4 & 0.9573 & $1.83 \times 10^{-3}$ & 46 & 36 & 0.0055 & -- & 0.0034 & 25.5 & 10.9 & 0.01 & 0.1 \\
 & 5 & 0.9960 & $1.63 \times 10^{-5}$ & 56 & 45 & 0.0055 & -- & 0.0054 & 28.7 & 12.9 & 0.01 & 0.2 \\
\midrule
MPD & 1 & 0.9804 & $3.83 \times 10^{-4}$ & 82 & 23 & 0.0055 & -- & 0.0000 & -- & 0.4 & 0.01 & 0.1 \\
 & 2 & 0.9848 & $2.31 \times 10^{-4}$ & 102 & 48 & 0.0055 & -- & 0.0029 & -- & 0.4 & 0.01 & 0.1 \\
 & 3 & 0.9875 & $1.56 \times 10^{-4}$ & 124 & 73 & 0.0055 & -- & 0.0045 & -- & 0.4 & 0.01 & 0.1 \\
 & 4 & 0.9897 & $1.05 \times 10^{-4}$ & 149 & 93 & 0.0055 & -- & 0.0049 & -- & 0.4 & 0.01 & 0.1 \\
 & 5 & 0.9902 & $9.61 \times 10^{-5}$ & 170 & 116 & 0.0055 & -- & 0.0058 & -- & 0.5 & 0.01 & 0.2 \\
\midrule
Exact & -- & 1.0000 & $0$ & 1696 & 876 & 0.0055 & -- & 0.0055 & -- & 0.2 & 13.66 & 168.9 \\
\bottomrule
\end{tabular}}
\end{table}

\begin{table}[ht]
\centering
\caption{Full benchmark results for $n_q = 12$ (2D normal).}
\label{tab:appendix-2d-normal-12q}
\resizebox{\textwidth}{!}{%
\begin{tabular}{l r r r r r r r r r r r r r}
\toprule
Method & $L$ & Accuracy & Infidelity & Depth & CX & $S_\mathrm{init}$ & $S_\mathrm{red}$ & $S_\mathrm{final}$ & Train (s) & Mem (MB) & Tr.\ time (s) & Tr.\ mem (MB) \\
\midrule
VDSP & 1 & 0.9800 & $4.00 \times 10^{-4}$ & 101 & 29 & 0.0058 & 0.0058 & 0.0000 & 21.8 & 19.8 & 0.01 & 0.1 \\
 & 2 & 0.9915 & $7.20 \times 10^{-5}$ & 118 & 50 & 0.0058 & 0.0020 & 0.0054 & 31.6 & 11.7 & 0.01 & 0.1 \\
 & 3 & 0.9944 & $3.17 \times 10^{-5}$ & 131 & 63 & 0.0058 & 0.0009 & 0.0061 & 45.1 & 13.9 & 0.01 & 0.1 \\
 & 4 & 0.9950 & $2.48 \times 10^{-5}$ & 136 & 74 & 0.0058 & 0.0012 & 0.0060 & 68.3 & 16.9 & 0.01 & 0.1 \\
 & 5 & 0.9972 & $7.58 \times 10^{-6}$ & 146 & 83 & 0.0058 & 0.0004 & 0.0055 & 65.3 & 19.1 & 0.01 & 0.1 \\
\midrule
PQC & 1 & 0.7870 & $4.48 \times 10^{-2}$ & 16 & 11 & 0.0058 & -- & 0.0000 & 12.9 & 6.1 & 0.01 & 0.1 \\
 & 2 & 0.9330 & $4.49 \times 10^{-3}$ & 26 & 22 & 0.0058 & -- & 0.0058 & 18.9 & 8.2 & 0.01 & 0.1 \\
 & 3 & 0.9540 & $2.12 \times 10^{-3}$ & 36 & 33 & 0.0058 & -- & 0.0039 & 25.3 & 10.3 & 0.01 & 0.1 \\
 & 4 & 0.9569 & $1.86 \times 10^{-3}$ & 46 & 44 & 0.0058 & -- & 0.0041 & 28.1 & 13.0 & 0.01 & 0.1 \\
 & 5 & 0.9905 & $9.06 \times 10^{-5}$ & 56 & 55 & 0.0058 & -- & 0.0063 & 35.0 & 15.6 & 0.01 & 0.1 \\
\midrule
MPD & 1 & 0.9800 & $4.00 \times 10^{-4}$ & 98 & 28 & 0.0058 & -- & 0.0000 & -- & 1.4 & 0.01 & 0.1 \\
 & 2 & 0.9845 & $2.41 \times 10^{-4}$ & 122 & 60 & 0.0058 & -- & 0.0035 & -- & 1.4 & 0.01 & 0.1 \\
 & 3 & 0.9858 & $2.02 \times 10^{-4}$ & 143 & 86 & 0.0058 & -- & 0.0042 & -- & 1.4 & 0.01 & 0.1 \\
 & 4 & 0.9886 & $1.30 \times 10^{-4}$ & 165 & 113 & 0.0058 & -- & 0.0058 & -- & 1.4 & 0.02 & 0.1 \\
 & 5 & 0.9890 & $1.20 \times 10^{-4}$ & 184 & 139 & 0.0058 & -- & 0.0055 & -- & 1.4 & 0.01 & 0.1 \\
\midrule
Exact & -- & 1.0000 & $0$ & 6695 & 3613 & 0.0058 & -- & 0.0058 & -- & 0.5 & 0.04 & 0.4 \\
\bottomrule
\end{tabular}}
\end{table}

\begin{table}[ht]
\centering
\caption{Full benchmark results for $n_q = 14$ (2D normal).}
\label{tab:appendix-2d-normal-14q}
\resizebox{\textwidth}{!}{%
\begin{tabular}{l r r r r r r r r r r r r r}
\toprule
Method & $L$ & Accuracy & Infidelity & Depth & CX & $S_\mathrm{init}$ & $S_\mathrm{red}$ & $S_\mathrm{final}$ & Train (s) & Mem (MB) & Tr.\ time (s) & Tr.\ mem (MB) \\
\midrule
VDSP & 1 & 0.9799 & $4.04 \times 10^{-4}$ & 116 & 34 & 0.0060 & 0.0060 & 0.0000 & 47.0 & 23.8 & 0.01 & 0.1 \\
 & 2 & 0.9905 & $9.02 \times 10^{-5}$ & 135 & 61 & 0.0060 & 0.0027 & 0.0056 & 76.1 & 15.0 & 0.01 & 0.1 \\
 & 3 & 0.9945 & $3.07 \times 10^{-5}$ & 148 & 74 & 0.0060 & 0.0011 & 0.0053 & 94.5 & 17.5 & 0.02 & 0.2 \\
 & 4 & 0.9935 & $4.19 \times 10^{-5}$ & 157 & 87 & 0.0060 & 0.0015 & 0.0052 & 109.1 & 22.7 & 17.84 & 169.1 \\
 & 5 & 0.9933 & $4.49 \times 10^{-5}$ & 163 & 94 & 0.0060 & 0.0011 & 0.0058 & 241.7 & 22.9 & 0.02 & 0.1 \\
\midrule
PQC & 1 & 0.7871 & $4.48 \times 10^{-2}$ & 16 & 13 & 0.0060 & -- & 0.0000 & 14.6 & 6.8 & 0.01 & 0.1 \\
 & 2 & 0.9322 & $4.59 \times 10^{-3}$ & 26 & 26 & 0.0060 & -- & 0.0042 & 27.5 & 10.0 & 0.01 & 0.1 \\
 & 3 & 0.9528 & $2.23 \times 10^{-3}$ & 36 & 39 & 0.0060 & -- & 0.0034 & 27.6 & 12.3 & 0.01 & 0.2 \\
 & 4 & 0.9528 & $2.23 \times 10^{-3}$ & 46 & 52 & 0.0060 & -- & 0.0033 & 34.0 & 15.0 & 0.01 & 0.2 \\
 & 5 & 0.9573 & $1.82 \times 10^{-3}$ & 56 & 65 & 0.0060 & -- & 0.0039 & 42.9 & 18.4 & 0.01 & 0.1 \\
\midrule
MPD & 1 & 0.9799 & $4.04 \times 10^{-4}$ & 114 & 29 & 0.0060 & -- & 0.0000 & -- & 5.6 & 0.01 & 0.1 \\
 & 2 & 0.9861 & $1.94 \times 10^{-4}$ & 138 & 64 & 0.0060 & -- & 0.0038 & -- & 5.6 & 0.01 & 0.1 \\
 & 3 & 0.9868 & $1.75 \times 10^{-4}$ & 157 & 92 & 0.0060 & -- & 0.0066 & -- & 5.6 & 0.01 & 0.1 \\
 & 4 & 0.9889 & $1.22 \times 10^{-4}$ & 179 & 133 & 0.0060 & -- & 0.0070 & -- & 5.6 & 0.02 & 0.2 \\
 & 5 & 0.9886 & $1.31 \times 10^{-4}$ & 207 & 172 & 0.0060 & -- & 0.0064 & -- & 5.6 & 0.02 & 0.1 \\
\midrule
Exact & -- & 1.0000 & $0$ & 27051 & 14761 & 0.0060 & -- & 0.0060 & -- & 1.8 & 0.18 & 1.5 \\
\bottomrule
\end{tabular}}
\end{table}

\begin{table}[ht]
\centering
\caption{Full benchmark results for $n_q = 16$ (2D normal).}
\label{tab:appendix-2d-normal-16q}
\resizebox{\textwidth}{!}{%
\begin{tabular}{l r r r r r r r r r r r r r}
\toprule
Method & $L$ & Accuracy & Infidelity & Depth & CX & $S_\mathrm{init}$ & $S_\mathrm{red}$ & $S_\mathrm{final}$ & Train (s) & Mem (MB) & Tr.\ time (s) & Tr.\ mem (MB) \\
\midrule
VDSP & 1 & 0.9799 & $4.06 \times 10^{-4}$ & 131 & 36 & 0.0060 & 0.0060 & 0.0000 & 132.2 & 32.9 & 0.01 & 0.1 \\
 & 2 & 0.9924 & $5.79 \times 10^{-5}$ & 154 & 67 & 0.0060 & 0.0014 & 0.0044 & 180.4 & 21.6 & 0.01 & 0.1 \\
 & 3 & 0.9936 & $4.13 \times 10^{-5}$ & 160 & 83 & 0.0060 & 0.0017 & 0.0061 & 190.1 & 24.2 & 0.01 & 0.1 \\
 & 4 & 0.9943 & $3.26 \times 10^{-5}$ & 176 & 99 & 0.0060 & 0.0009 & 0.0051 & 454.8 & 28.7 & 0.01 & 0.1 \\
 & 5 & 0.9944 & $3.13 \times 10^{-5}$ & 182 & 114 & 0.0060 & 0.0010 & 0.0056 & 379.7 & 31.4 & 17.99 & 176.4 \\
\midrule
PQC & 1 & 0.7866 & $4.50 \times 10^{-2}$ & 16 & 15 & 0.0060 & -- & 0.0000 & 17.3 & 8.0 & 0.01 & 0.1 \\
 & 2 & 0.9317 & $4.66 \times 10^{-3}$ & 26 & 30 & 0.0060 & -- & 0.0058 & 25.5 & 10.8 & 0.01 & 0.1 \\
 & 3 & 0.9498 & $2.52 \times 10^{-3}$ & 36 & 45 & 0.0060 & -- & 0.0033 & 34.5 & 14.6 & 0.01 & 0.1 \\
 & 4 & 0.9568 & $1.86 \times 10^{-3}$ & 46 & 60 & 0.0060 & -- & 0.0036 & 40.7 & 18.1 & 0.01 & 0.1 \\
 & 5 & 0.9569 & $1.86 \times 10^{-3}$ & 56 & 75 & 0.0060 & -- & 0.0038 & 48.3 & 21.8 & 0.01 & 0.2 \\
\midrule
MPD & 1 & 0.9799 & $4.06 \times 10^{-4}$ & 133 & 38 & 0.0060 & -- & 0.0000 & -- & 11.4 & 0.01 & 0.1 \\
 & 2 & 0.9871 & $1.68 \times 10^{-4}$ & 150 & 73 & 0.0060 & -- & 0.0042 & -- & 11.9 & 0.01 & 0.1 \\
 & 3 & 0.9856 & $2.07 \times 10^{-4}$ & 175 & 114 & 0.0060 & -- & 0.0042 & -- & 11.9 & 0.02 & 0.1 \\
 & 4 & 0.9870 & $1.68 \times 10^{-4}$ & 199 & 154 & 0.0060 & -- & 0.0065 & -- & 12.0 & 0.02 & 0.1 \\
 & 5 & 0.9889 & $1.23 \times 10^{-4}$ & 218 & 189 & 0.0060 & -- & 0.0063 & -- & 12.0 & 0.02 & 0.2 \\
\midrule
Exact & -- & 1.0000 & $0$ & 108149 & 59517 & 0.0060 & -- & 0.0060 & -- & 7.1 & 0.94 & 6.1 \\
\bottomrule
\end{tabular}}
\end{table}

\begin{table}[ht]
\centering
\caption{Full benchmark results for $n_q = 18$ (2D normal).}
\label{tab:appendix-2d-normal-18q}
\resizebox{\textwidth}{!}{%
\begin{tabular}{l r r r r r r r r r r r r r}
\toprule
Method & $L$ & Accuracy & Infidelity & Depth & CX & $S_\mathrm{init}$ & $S_\mathrm{red}$ & $S_\mathrm{final}$ & Train (s) & Mem (MB) & Tr.\ time (s) & Tr.\ mem (MB) \\
\midrule
VDSP & 1 & 0.9799 & $4.06 \times 10^{-4}$ & 147 & 42 & 0.0061 & 0.0061 & 0.0000 & 491.4 & 61.8 & 0.02 & 0.1 \\
 & 2 & 0.9925 & $5.58 \times 10^{-5}$ & 171 & 82 & 0.0061 & 0.0016 & 0.0044 & 746.2 & 46.4 & 0.02 & 0.1 \\
 & 3 & 0.9928 & $5.22 \times 10^{-5}$ & 182 & 94 & 0.0061 & 0.0022 & 0.0050 & 791.9 & 52.0 & 0.02 & 0.7 \\
 & 4 & 0.9926 & $5.45 \times 10^{-5}$ & 192 & 112 & 0.0061 & 0.0017 & 0.0049 & 857.0 & 51.4 & 0.01 & 0.1 \\
 & 5 & 0.9944 & $3.15 \times 10^{-5}$ & 203 & 128 & 0.0061 & 0.0012 & 0.0053 & 910.1 & 54.4 & 0.02 & 0.1 \\
\midrule
PQC & 1 & 0.7868 & $4.49 \times 10^{-2}$ & 16 & 17 & 0.0061 & -- & 0.0000 & 33.0 & 12.9 & 0.01 & 0.1 \\
 & 2 & 0.9312 & $4.73 \times 10^{-3}$ & 26 & 34 & 0.0061 & -- & 0.0029 & 35.7 & 15.7 & 0.01 & 0.1 \\
 & 3 & 0.9521 & $2.30 \times 10^{-3}$ & 36 & 51 & 0.0061 & -- & 0.0033 & 41.8 & 19.2 & 0.01 & 0.2 \\
 & 4 & 0.9560 & $1.94 \times 10^{-3}$ & 46 & 68 & 0.0061 & -- & 0.0036 & 53.6 & 23.3 & 0.01 & 0.1 \\
 & 5 & 0.9569 & $1.86 \times 10^{-3}$ & 56 & 85 & 0.0061 & -- & 0.0038 & 69.9 & 27.7 & 0.01 & 0.1 \\
\midrule
MPD & 1 & 0.9799 & $4.06 \times 10^{-4}$ & 147 & 45 & 0.0061 & -- & 0.0000 & -- & 30.0 & 0.01 & 0.1 \\
 & 2 & 0.9839 & $2.58 \times 10^{-4}$ & 169 & 87 & 0.0061 & -- & 0.0032 & -- & 30.0 & 0.01 & 0.1 \\
 & 3 & 0.9843 & $2.46 \times 10^{-4}$ & 192 & 125 & 0.0061 & -- & 0.0044 & -- & 30.1 & 0.01 & 0.2 \\
 & 4 & 0.9881 & $1.41 \times 10^{-4}$ & 210 & 167 & 0.0061 & -- & 0.0059 & -- & 30.1 & 0.02 & 0.1 \\
 & 5 & 0.9913 & $7.55 \times 10^{-5}$ & 237 & 217 & 0.0061 & -- & 0.0069 & -- & 30.2 & 0.02 & 0.1 \\
\midrule
Exact & -- & 1.0000 & $0$ & 433835 & 239304 & 0.0061 & -- & 0.0061 & -- & 28.1 & 23.37 & 176.7 \\
\bottomrule
\end{tabular}}
\end{table}

\begin{table}[ht]
\centering
\caption{Full benchmark results for $n_q = 20$ (2D normal).}
\label{tab:appendix-2d-normal-20q}
\resizebox{\textwidth}{!}{%
\begin{tabular}{l r r r r r r r r r r r r r}
\toprule
Method & $L$ & Accuracy & Infidelity & Depth & CX & $S_\mathrm{init}$ & $S_\mathrm{red}$ & $S_\mathrm{final}$ & Train (s) & Mem (MB) & Tr.\ time (s) & Tr.\ mem (MB) \\
\midrule
VDSP & 1 & 0.9799 & $4.06 \times 10^{-4}$ & 166 & 49 & 0.0061 & 0.0061 & 0.0000 & 1257.9 & 163.5 & 0.01 & 0.1 \\
 & 2 & 0.9922 & $6.01 \times 10^{-5}$ & 184 & 85 & 0.0061 & 0.0019 & 0.0043 & 1976.4 & 155.7 & 0.02 & 0.7 \\
 & 3 & 0.9949 & $2.56 \times 10^{-5}$ & 192 & 104 & 0.0061 & 0.0012 & 0.0060 & 2190.2 & 150.8 & 0.02 & 0.7 \\
 & 4 & 0.9953 & $2.22 \times 10^{-5}$ & 204 & 125 & 0.0061 & 0.0011 & 0.0055 & 2606.8 & 142.1 & 0.02 & 0.7 \\
 & 5 & 0.9961 & $1.54 \times 10^{-5}$ & 217 & 143 & 0.0061 & 0.0008 & 0.0056 & 4788.2 & 142.1 & 0.02 & 0.1 \\
\midrule
PQC & 1 & 0.7866 & $4.50 \times 10^{-2}$ & 16 & 19 & 0.0061 & -- & 0.0000 & 50.4 & 37.4 & 0.01 & 0.1 \\
 & 2 & 0.9064 & $8.74 \times 10^{-3}$ & 26 & 38 & 0.0061 & -- & 0.0032 & 124.5 & 40.6 & 0.01 & 0.2 \\
 & 3 & 0.9530 & $2.20 \times 10^{-3}$ & 36 & 57 & 0.0061 & -- & 0.0037 & 79.9 & 43.4 & 0.01 & 0.2 \\
 & 4 & 0.9568 & $1.86 \times 10^{-3}$ & 46 & 76 & 0.0061 & -- & 0.0038 & 91.0 & 46.7 & 0.01 & 0.2 \\
 & 5 & 0.9568 & $1.86 \times 10^{-3}$ & 56 & 95 & 0.0061 & -- & 0.0040 & 168.7 & 49.8 & 0.01 & 0.1 \\
\midrule
MPD & 1 & 0.9799 & $4.06 \times 10^{-4}$ & 165 & 46 & 0.0061 & -- & 0.0000 & -- & 93.7 & 0.01 & 0.1 \\
 & 2 & 0.9853 & $2.17 \times 10^{-4}$ & 186 & 97 & 0.0061 & -- & 0.0047 & -- & 93.8 & 0.01 & 0.2 \\
 & 3 & 0.9847 & $2.35 \times 10^{-4}$ & 214 & 151 & 0.0061 & -- & 0.0049 & -- & 93.9 & 0.02 & 0.2 \\
 & 4 & 0.9870 & $1.68 \times 10^{-4}$ & 229 & 188 & 0.0061 & -- & 0.0076 & -- & 94.0 & 0.02 & 0.2 \\
 & 5 & 0.9867 & $1.76 \times 10^{-4}$ & 251 & 245 & 0.0061 & -- & 0.0061 & -- & 94.0 & 0.02 & 0.1 \\
\midrule
Exact & -- & 1.0000 & $0$ & 1735314 & 959010 & 0.0061 & -- & 0.0061 & -- & 112.1 & 53.43 & 200.3 \\
\bottomrule
\end{tabular}}
\end{table}

\end{document}